\RequirePackage{mathtools}
\documentclass[online]{jfp1}

\usepackage{hyperref}
\usepackage[all]{xy}

\usepackage{stmaryrd}
\usepackage{ucs}
\usepackage[utf8x]{inputenc}
\usepackage{autofe}
\usepackage{proof}
\usepackage{framed}

\usepackage[usenames,dvipsnames]{color}
\usepackage{soul}

\DeclareMathAlphabet{\mathkw}{OT1}{cmss}{bx}{n} 

\numberwithin{equation}{section}

\usepackage{ifthen}
\newboolean{ExtendedVersion}
\setboolean{ExtendedVersion}{false}     

\newcommand{\extended}[2][{}]{\ifthenelse{\boolean{ExtendedVersion}}{#2}{#1}}

\def\myproof{\noindent{\em Proof\/}}

\def\squareforqed{\hbox{\rlap{$\sqcap$}$\sqcup$}}

\def\qed{\ifmmode\squareforqed\else{\unskip\nobreak\hfil
\penalty50\hskip1em\null\nobreak\hfil\squareforqed
\parfillskip=0pt\finalhyphendemerits=0\endgraf}\fi}

\newcommand{\myqed}{\qed\vspace*{5pt}}

\newenvironment{mproof}{\myproof}{\myqed}

\newlength{\nl}
\newtheorem{mtheorem}{Theorem}[section]
\newtheorem{mprop}[mtheorem]{Proposition}
\newtheorem{mdefinition}[mtheorem]{Definition}
\newtheorem{mlemma}[mtheorem]{Lemma}

\newtheorem{mremark}[mtheorem]{Remark}

\newcommand{\FN}{\ensuremath\mathsf} 

\newcommand{\Naturals}{\mathbb{N}}  
\newcommand{\Set}{\mathbf{\mathsf{Set}}} 
\newcommand{\cat}[1]{\mathcal{#1}} 
\newcommand{\C}{\cat{C}}  
\newcommand{\D}{\cat{D}}  

\newcommand{\op}{\FN{op}}

\newcommand{\EndD}{\FN{End_{\dayt}}}

\newcommand{\EndC}{\FN{End_{\circ}}}
\newcommand{\dayt}{\mathrel{\star}}

\newcommand{\id}{\FN{id}} 
\newcommand{\Id}{\FN{Id}}
\newcommand{\inl}{\FN{inl}} 
\newcommand{\inr}{\FN{inr}} 
\newcommand{\oX}{\mathrel{\otimes}}
\newcommand{\oI}{I}
\newcommand{\oP}{\mathrel{\oplus}}
\newcommand{\oJ}{J}
\newcommand{\Monoid}[1]{\mathsf{Mon({#1})}} 

\newcommand{\rep}{\FN{rep}}
\newcommand{\abs}{\FN{abs}}

\newcommand{\free}[1]{\FN{free}\ {#1}} 
\newcommand{\ins}{\FN{ins}} 

\newcommand{\Rexp}[2]{{#2}^{#1}}

\newcommand{\ladj}[1]{\lfloor {#1} \rfloor} 
\newcommand{\radj}[1]{\lceil {#1} \rceil} 
\newcommand{\ev}{\mathsf{ev}} 

\newcommand{\Prof}[2]{\mathsf{Prof}({#1}, {#2})}
\newcommand{\Pro}{\mathsf{Pro}}
\newcommand{\SPro}{\mathsf{SPro}}
\newcommand{\Endo}{{\mathsf{End}_\circ}} 
\newcommand{\Hom}{\FN{Hom}}

\newcommand{\profunctor}{\mathrel{\longrightarrow\!\!\!\!\!\!\!\!\!\!\!\!+\,\,\,\,}}

\newcommand{\Dinat}[2]{\FN{Dinat}({#1},{#2})}
\newcommand{\Nat}[2]{\FN{Nat}({#1},{#2})}

\newenvironment{coder}{\begin{framed}}{\end{framed}}

\newcommand{\exponentiable}{an exponent}
\newcommand{\verylongrightarrow}{\xrightarrow{\hspace*{1cm}}}

\newcommand{\comment}[1]{\quad\{ \text{ #1 } \}}
\newcommand{\banana}[1]{\llparenthesis {#1}\rrparenthesis}
\newcommand{\str}{\FN{str}}
\newcommand{\sst}{\FN{st}}
\newcommand{\funst}{\FN{st}}

\makeatletter
\@ifundefined{lhs2tex.lhs2tex.sty.read}%
  {\@namedef{lhs2tex.lhs2tex.sty.read}{}%
   \newcommand\SkipToFmtEnd{}%
   \newcommand\EndFmtInput{}%
   \long\def\SkipToFmtEnd#1\EndFmtInput{}%
  }\SkipToFmtEnd

\newcommand\ReadOnlyOnce[1]{\@ifundefined{#1}{\@namedef{#1}{}}\SkipToFmtEnd}
\usepackage{amstext}
\usepackage{amssymb}
\usepackage{stmaryrd}
\DeclareFontFamily{OT1}{cmtex}{}
\DeclareFontShape{OT1}{cmtex}{m}{n}
  {<5><6><7><8>cmtex8
   <9>cmtex9
   <10><10.95><12><14.4><17.28><20.74><24.88>cmtex10}{}
\DeclareFontShape{OT1}{cmtex}{m}{it}
  {<-> ssub * cmtt/m/it}{}

\DeclareFontShape{OT1}{cmtt}{bx}{n}
  {<5><6><7><8>cmtt8
   <9>cmbtt9
   <10><10.95><12><14.4><17.28><20.74><24.88>cmbtt10}{}
\DeclareFontShape{OT1}{cmtex}{bx}{n}
  {<-> ssub * cmtt/bx/n}{}

\newcommand{\Varid}[1]{\mathit{#1}}
\newcommand{\anonymous}{\kern0.06em \vbox{\hrule\@width.5em}}
\newcommand{\plus}{\mathbin{+\!\!\!+}}
\newcommand{\bind}{\mathbin{>\!\!\!>\mkern-6.7mu=}}

\usepackage{polytable}

\@ifundefined{mathindent}%
  {\newdimen\mathindent\mathindent\leftmargini}%
  {}%

\def\resethooks{%
  \global\let\SaveRestoreHook\empty
  \global\let\ColumnHook\empty}
\newcommand*{\savecolumns}[1][default]%
  {\g@addto@macro\SaveRestoreHook{\savecolumns[#1]}}
\newcommand*{\restorecolumns}[1][default]%
  {\g@addto@macro\SaveRestoreHook{\restorecolumns[#1]}}
\newcommand*{\aligncolumn}[2]%
  {\g@addto@macro\ColumnHook{\column{#1}{#2}}}

\resethooks

\newcommand{\onelinecommentchars}{\quad-{}- }
\newcommand{\commentbeginchars}{\enskip\{-}
\newcommand{\commentendchars}{-\}\enskip}

\newcommand{\visiblecomments}{%
  \let\onelinecomment=\onelinecommentchars
  \let\commentbegin=\commentbeginchars
  \let\commentend=\commentendchars}

\newcommand{\invisiblecomments}{%
  \let\onelinecomment=\empty
  \let\commentbegin=\empty
  \let\commentend=\empty}

\visiblecomments

\newlength{\blanklineskip}
\newcommand{\hsindent}[1]{\quad}
\let\hspre\empty
\let\hspost\empty

\EndFmtInput
\makeatother
\ReadOnlyOnce{polycode.fmt}%
\makeatletter

\newcommand{\hsnewpar}[1]%
  {{\parskip=0pt\parindent=0pt\par\vskip #1\noindent}}

\newcommand{\hscodestyle}{}

\newcommand{\sethscode}[1]%
  {\expandafter\let\expandafter\hscode\csname #1\endcsname
   \expandafter\let\expandafter\endhscode\csname end#1\endcsname}

  {\par\noindent
   \advance\leftskip\mathindent
   \hscodestyle
   \let\\=\@normalcr
   \let\hspre\(\let\hspost\)%
   \pboxed}%
  {\endpboxed\)%
   \par\noindent
   \ignorespacesafterend}

  {\hsnewpar\abovedisplayskip
   \advance\leftskip\mathindent
   \hscodestyle
   \let\hspre\(\let\hspost\)%
   \pboxed}%
  {\endpboxed%
   \hsnewpar\belowdisplayskip
   \ignorespacesafterend}

  {\hsnewpar\abovedisplayskip
   \advance\leftskip\mathindent
   \hscodestyle
   \let\\=\@normalcr
   \(\pboxed}%
  {\endpboxed\)%
   \hsnewpar\belowdisplayskip
   \ignorespacesafterend}

\newcommand{\plainhs}{\sethscode{plainhscode}}

\plainhs

  {\hsnewpar\abovedisplayskip
   \advance\leftskip\mathindent
   \hscodestyle
   \let\\=\@normalcr
   \(\parray}%
  {\endparray\)%
   \hsnewpar\belowdisplayskip
   \ignorespacesafterend}

  {\parray}{\endparray}

  {\(\parray}{\endparray\)}

\def\codeframewidth{\arrayrulewidth}
\RequirePackage{calc}

  {\parskip=\abovedisplayskip\par\noindent
   \hscodestyle
   \arrayrulewidth=\codeframewidth
   \tabular{@{}|p{\linewidth-2\arraycolsep-2\arrayrulewidth-2pt}|@{}}%
   \hline\framedhslinecorrect\\{-1.5ex}%
   \let\endoflinesave=\\
   \let\\=\@normalcr
   \(\pboxed}%
  {\endpboxed\)%
   \framedhslinecorrect\endoflinesave{.5ex}\hline
   \endtabular
   \parskip=\belowdisplayskip\par\noindent
   \ignorespacesafterend}

\newcommand{\framedhslinecorrect}[2]%
  {#1[#2]}

  {\(\def\column##1##2{}%
   \let\>\undefined\let\<\undefined\let\\\undefined
   \newcommand\>[1][]{}\newcommand\<[1][]{}\newcommand\\[1][]{}%
   \def\fromto##1##2##3{##3}%
   }{\) }%

  {\let\orighscode=\hscode
   \let\origendhscode=\endhscode
   \def\endhscode{\def\hscode{\endgroup\def\@currenvir{hscode}\\}\begingroup}
   \orighscode\def\hscode{\endgroup\def\@currenvir{hscode}}}%
  {\origendhscode
   \global\let\hscode=\orighscode
   \global\let\endhscode=\origendhscode}%

\makeatother
\EndFmtInput
\ReadOnlyOnce{forall.fmt}%
\makeatletter

\let\HaskellResetHook\empty
\newcommand*{\AtHaskellReset}[1]{%
  \g@addto@macro\HaskellResetHook{#1}}
\newcommand*{\HaskellReset}{\HaskellResetHook}

\newcommand\hsforall{\global\let\hsdot=\hsperiodonce}
\newcommand*\hsperiodonce[2]{#2\global\let\hsdot=\hscompose}
\newcommand*\hscompose[2]{#1}

\AtHaskellReset{\global\let\hsdot=\hscompose}

\HaskellReset

\makeatother
\EndFmtInput

\newcommand{\sbt}{\begin{picture}(-1,-1)\circle*{1.5}\end{picture}\,}

\title[Notions of Computation as Monoids]
    {Notions of Computation as Monoids}
\author[E. Rivas and M.\ Jaskelioff]
   { EXEQUIEL RIVAS\qquad MAURO JASKELIOFF\\
    Centro Internacional Franco Argentino de Ciencias de la Informaci\'on y de Sistemas\\
    CONICET, Argentina \qquad\quad
    FCEIA, Universidad Nacional de Rosario, Argentina\\[8pt] 
    }{}

\begin{document}
\maketitle

\begin{abstract}
  There are different notions of computation, the most popular being
  monads, applicative functors, and arrows. In this article we show
  that these three notions can be seen as monoids in a monoidal
  category.  We demonstrate that at this level of abstraction one can
  obtain useful results which can be instantiated to the different
  notions of computation. In particular, we show how free
  constructions and Cayley representations for monoids translate into
  useful constructions for monads, applicative functors, and
  arrows. Moreover, the uniform presentation of all three notions
  helps in the analysis of the relation between them.
\end{abstract}

\section{Introduction}
\label{sec:intro}

When constructing a semantic model of a system or when structuring
computer code, there are several notions of computation that one might
consider. Monads~\cite{Moggi89,Moggi91} are the most popular notion, but other
notions, such as arrows~\cite{Hughes-SCP00} and, more recently,
applicative functors~\cite{mcbride08:applicative-programming} have
been gaining widespread acceptance.

Each of these notions of computation has particular characteristics that
makes them more suitable for some tasks than for others. 
Nevertheless, there is much to be gained from unifying all three different
notions under a single conceptual framework.

In this article we show how all three of these notions of computation
can be cast as a monoid in a monoidal category. Monads are known to be
monoids in a monoidal category of
endofunctors~\cite{macLaneS:catwm,ttt}. Moreover, strong monads are
monoids in a monoidal category of strong endofunctors.  Arrows have
been recently shown to be strong monoids in a monoidal category of
profunctors by Jacobs et al.~\shortcite{jacobs2009categorical}.
Applicative functors, on the other hand, are usually presented as lax
monoidal functors with a compatible
strength~\cite{mcbride08:applicative-programming,Jaskelioff:MSFP2012,PatersonMPC}.
However, in the category-theory community, it is known that lax
monoidal functors are monoids with respect to the Day convolution, and
hence applicative functors are also monoids in a monoidal category of
endofunctors using the Day convolution as a tensor~\cite{GoodDay}.

Therefore, we unify the analysis of three different notions of
computation, namely monads, applicative functors, and arrows, by
looking at them as monoids in a monoidal category. 
In particular, we make explicit the relation between applicative
functors and monoids with respect to the Day convolution, and we
simplify the characterisation of arrows.
Unlike the approach to arrows of Jacobs et
al.~\shortcite{jacobs2009categorical}, where the operation \ensuremath{\FN{first}} is added on
top of the monoid structure, we obtain that operation from the monoidal
structure of the underlying category.  Furthermore, we show that at
the level of abstraction of monoidal categories one can obtain useful
results, such as free constructions and Cayley representations.

Free constructions are often used in programming in order to represent
abstract syntax trees. For instance, free constructions are used to
define deep embeddings of domain-specific languages. Traditionally,
one uses a free monad to represent abstract syntax trees, with the
bind operation (Kleisli extension) acting as a form of simultaneous
substitution. However, in certain cases, the free applicative functor
is a better fit~\cite{Capriotti2014}. The free arrow, on the other
hand, has been less explored and we know of no publication that has an
implementation of it.

The Cayley representation theorem states that every group is
isomorphic to a group of permutations~\cite{Cayley}. Hence, one can
work with a concrete group of permutations instead of working with an
abstract group. The representation theorem does not really use the
inverse operation of groups so one can generalise the representation to
monoids, yielding a Cayley representation theorem for
monoids~\cite{jacobson2009basic}.

In functional programming, the Cayley theorem appears as an
optimisation by change of representation. We identify two known
optimisations, namely difference lists~\cite{Hughes86}, and the
codensity monad transformation~\cite{Voigtlander08,F5} as being
essentially the same, since both are instances of the general Cayley
representation of monoids in a monoidal category. Moreover, we obtain
similar transformations for applicative functors and arrows by
analysing their Cayley representations.

Given the three notions of computation, one may ask what is the
relation between them. 
Lindley et al.~\shortcite{LindleyWY11} address this question by
studying the equational theories induced by each calculus.
Since the different notions are monoids in a monoidal category, a categorical approach could be to ask about the relation between the
corresponding categories of monoids. However, another consequence of
having a unified view is that we can ask a deeper question instead and
analyse the relation between the different monoidal categories that
support them. Then, we obtain the relation between their monoids as a
corollary.

Concretely, the article makes the following contributions:
\begin{itemize}
\item We present a unified view of monads, applicative functors, and
  arrows as monoids in a monoidal category. Although the results are
  known in other communities, the case of the applicative functors as
  monoids seems to have been overlooked in the functional programming
  community.
\item We show how the Cayley representation of monoids unifies two
  different known optimisations, namely difference lists and the
  codensity monad transformation. The similarity between these two
  optimisations has been noticed before, but now we make the relation
  precise and demonstrate that they are two instances of the
  \emph{same} change of representation.
\item We apply the characterisation of applicative functors as monoids
  to obtain a free construction and a Cayley representation for
  applicative functors. In this way, we clarify the construction of
  free applicative functors as explained by Capriotti and
  Kaposi~\shortcite{Capriotti2014}. The Cayley representation for
  applicative functors is entirely new.
\item We clarify the view of arrows as monoids by introducing the
  strength in the monoidal category. In previous approaches, the
  strength was added to the monoids, while in this article we consider
  a category with strong profunctors. Our approach leads to a new
  categorical model of arrows and to the first formulation of free
  arrows.
\item We analyse the relation between the monoidal categories that
  give rise to monads, applicative functors, and arrows, by
  constructing monoidal functors between them.
\end{itemize}

The rest of the article is structured as follows. 
In Section~\ref{sec:CayleySet} we introduce the
Cayley representation for ordinary monoids. 
In Section~\ref{sec:monoidalcat}, we introduce monoidal categories,
monoids, free monoids and the Cayley representation for monoids in a
monoidal category.
In Section~\ref{sec:monads}, we instantiate these constructions to a
category of endofunctors, with composition as a tensor and obtain
monads, free monads, and the Cayley representation for monads.
In Section~\ref{sec:applicative}, we do the same for applicative
functors. Before that, we introduce in Section~\ref{sec:ends} the
notions of ends and coends needed to define and work with the Day
convolution.
In Section~\ref{sec:prearrows}, we work in a category of Profunctors
to obtain pre-arrows, their free constructions, and their Cayley
representations.
In section~\ref{sec:arrows}, we turn to arrows, analyse the relation
between arrows and pre-arrows, and construct free arrows and an arrow
representation.
Finally, in section~\ref{sec:functors}, we analyse the relation between
the different monoidal categories considered in the previous sections,
and
conclude in Section~\ref{sec:conclusion}. 

The article is aimed at functional programmers with knowledge of basic
category theory concepts, such as categories, functors, and
adjunctions. We provide an introduction to more advanced concepts,
such as monoidal categories and ends.

\begin{coder}
  In frames like the one surrounding this paragraph, we include
  Haskell implementations of several of the categorical concepts of
  the article. The idea is not to formalise these concepts in Haskell,
  but rather to show how the category theory informs and guides the
  implementation.
\end{coder}

\subsection{Cayley representation for monoids}
\label{sec:CayleySet}

We start by stating the Cayley representation theorem for ordinary
monoids, i.e. monoids in the category of sets and functions $\Set$.  A
monoid is a triple $(M,\oplus,e)$ of a set $M$, a binary operation
$\oplus : M\times M \to M$ which is associative ($(a\oplus b) \oplus c
= a \oplus (b \oplus c)$), and an element $e\in M$ which is a left and
right identity with respect to the binary operation (i.e. $e \oplus a
= a = a \oplus e$.) Because of the obvious monoid
$(\Naturals,\cdot,1)$, the element $e$ and the binary operation
$\oplus$ are often called the \emph{unit} and \emph{multiplication} of
the monoid.

For every set $M$ we may construct the \emph {monoid of endomorphisms}
$(M \to M, \circ,\id)$, where $\circ$ is function composition and
$\id$ is the identity function.  

Up to an isomorphism, $M$ is a \emph{sub-monoid} of a monoid
$(M',\oplus',e')$ if there is an injection $i: M \hookrightarrow M'$
such that $ i(e) = e'$ and $i (a \oplus b) = i(a) \oplus' i(b)$ for
some $\oplus$ and $e$.
This makes $(M,\oplus,e)$ a monoid and $i$ a
\emph{monoid morphism}.

\begin{mtheorem}[Cayley representation for ($\Set$) monoids]
  \label{thm:Cayley_set}
  Every monoid $(M,\oplus,e)$ is a sub-monoid of the
  monoid of endomorphisms on $M$.
\end{mtheorem}
\begin{mproof}
  We construct an injection $\rep : M \to (M \to M)$ by currying the
  binary operation $\oplus$.
$$\rep(m) = \lambda m'.\, m\oplus  m'.$$%
The function $\rep$ is a monoid morphism:
\begin{align*}
\rep(e) & = \lambda m'.\, e\oplus m' \\[\nl]
     &  = \id\\
\rep (a \oplus b) & = \lambda m'.\, (a \oplus b)\oplus m' \\[\nl]
               & = \lambda m'.\, a \oplus (b \oplus m')\\[\nl]
               & = (\lambda m.\, a \oplus m) \circ (\lambda n.\, b \oplus n)\\[\nl] 
               & = \rep (a) \circ \rep(b)
\end{align*}
Moreover, $\rep$ is an injection, since we have a function $\abs : (M \to
  M) \to M$ given by 
  $$\abs(k) = k(e)$$
and, $\abs (\rep(m)) =  (\lambda m'.\, m\oplus m')\,e = m \oplus e = m$.

\end{mproof}

When $M$ lifts to a group (i.e. it has a compatible inverse
operation), then the monoid of endomorphisms on $M$ lifts to the
traditional Cayley representation of a group $M$.

\begin{coder}
 How can we use this theorem in Haskell? 
 Lists are monoids \ensuremath{([\mskip1.5mu \Varid{a}\mskip1.5mu],\plus ,[\mskip1.5mu \mskip1.5mu])} so we may apply
 Theorem~\ref{thm:Cayley_set}. Let us define a type synonym for the
 monoid of endomorphisms:
\begin{hscode}\SaveRestoreHook
\column{B}{@{}>{\hspre}l<{\hspost}@{}}%
\column{E}{@{}>{\hspre}l<{\hspost}@{}}%
\>[B]{}\mathkw{type}\;\mathsf{EList}\;\Varid{a}\mathrel{=}[\mskip1.5mu \Varid{a}\mskip1.5mu]\to [\mskip1.5mu \Varid{a}\mskip1.5mu]{}\<[E]%
\ColumnHook
\end{hscode}\resethooks
The functions \ensuremath{\rep} and \ensuremath{\abs} are
\begin{hscode}\SaveRestoreHook
\column{B}{@{}>{\hspre}l<{\hspost}@{}}%
\column{9}{@{}>{\hspre}c<{\hspost}@{}}%
\column{9E}{@{}l@{}}%
\column{13}{@{}>{\hspre}l<{\hspost}@{}}%
\column{E}{@{}>{\hspre}l<{\hspost}@{}}%
\>[B]{}\rep{}\<[9]%
\>[9]{}\mathbin{::}{}\<[9E]%
\>[13]{}[\mskip1.5mu \Varid{a}\mskip1.5mu]\to \mathsf{Elist}\;\Varid{a}{}\<[E]%
\\
\>[B]{}\rep\;\Varid{xs}{}\<[9]%
\>[9]{}\mathrel{=}{}\<[9E]%
\>[13]{}(\Varid{xs}\plus ){}\<[E]%
\\[\blanklineskip]%
\>[B]{}\abs{}\<[9]%
\>[9]{}\mathbin{::}{}\<[9E]%
\>[13]{}\mathsf{Elist}\;\Varid{a}\to [\mskip1.5mu \Varid{a}\mskip1.5mu]{}\<[E]%
\\
\>[B]{}\abs\;\Varid{xs}{}\<[9]%
\>[9]{}\mathrel{=}{}\<[9E]%
\>[13]{}\Varid{xs}\;[\mskip1.5mu \mskip1.5mu]{}\<[E]%
\ColumnHook
\end{hscode}\resethooks

By the theorem above, we have that \ensuremath{\abs\circ\rep\mathrel{=}\FN{id}}.
The type \ensuremath{\mathsf{Elist}\;\Varid{a}} is no other than difference lists!~\cite{Hughes86}.
Concatenation for standard lists is slow, as it
is linear on the first argument.
%
A well known solution is to use a different representation of lists:
the so-called ``difference lists'' or ``Hughes' lists'', in which
lists are represented by endofunctions of lists. 
%
%
%
%
In difference lists, concatenation is just function composition, and
the empty list is the identity function. Hence we can perform
efficient concatenations on difference lists, and when we are done we
can get back standard lists by applying the empty list.








\end{coder}

%

\section{Monoidal Categories}
\label{sec:monoidalcat}

The ordinary notion of monoid in the category $\Set$ of sets and
functions is too restrictive, so we are interested in generalising
monoids to other categories.
In order to express a monoid a category should have a notion of
\begin{enumerate}
\item a pairing operation for expressing the type of the multiplication,
\item and a type for expressing the unit.
\end{enumerate}

In $\Set$ (in fact, in any category with finite products), we may
define a binary operation on $X$ as a function $X \times X \to X$, and
the unit as a morphism $1 \to X$. However, a given category $\C$ may
not have finite products, or we may be interested in other monoidal
structure of $\C$, so we will be more general and abstract the product
by a $\otimes$ operation called a \emph{tensor}, and the unit $1$ by
an object $I$ of $\C$. Categories with a tensor $\oX$ and unit $I$
have the necessary structure for supporting an abstract notion of
monoid and are known as \emph{monoidal categories}.

\begin{mdefinition}[Monoidal Category]
A \emph{monoidal category} is a tuple $(\C,\oX,I,\alpha,\lambda,\rho)$, 
consisting of
\begin{itemize}
\item a category $\C$, 
\item a bifunctor $\oX : \C \times \C \to \C$,
\item an object $I$ of $\C$,
\item natural isomorphisms $\alpha_{A,B,C}:A\oX(B\oX C)\to (A\oX B)\oX C$ ,
  $\lambda_A:\oI\oX A\to A$ , and $\rho_A:A\oX\oI\to A$ such that 
  $\lambda_{\oI}=\rho_{\oI}$ and the following diagrams commute.
\[
\xymatrix{   
     A\oX(B\oX(C\oX D)) \ar[r]^{\alpha}
                                   \ar[d]_{\id\oX\alpha}
 &(A\oX B)\oX(C\oX D) \ar[r]^{\alpha}
&((A\oX B)\oX C)\oX D\\
A\oX((B\oX C)\oX D)\ar[rr]_{\alpha}&
&(A\oX(B\oX C))\oX D \ar[u]_{\alpha\oX\id}\\
}
\]
\[
\xymatrix{
A\oX(\oI\oX B)\ar[rr]^{\alpha}
\ar[dr]_{\id\oX\lambda} &
&(A\oX\oI)\oX B \ar[dl]^{\rho\oX\id}\\
& A\oX B\\
}\]
\end{itemize}

A monoidal category is said to be \emph{strict} when the natural
isomorphisms $\alpha$, $\lambda$ and $\rho$ are identities. Note that
in a strict monoidal category the diagrams necessarily commute.

A \emph{symmetric} monoidal category, is a monoidal category with an
additional natural isomorphism $\gamma_{A,B}: A\oX B \to B \oX A$
subject to some coherence conditions.
\end{mdefinition}

The idea of currying a function can be generalised to a monoidal
category with the following notion of exponential.

\begin{mdefinition}[Exponential]
\label{def:exp}
Let $A$ be an object of a monoidal category $(\C,\oX,I,\alpha,\lambda,\rho)$. An \emph{exponential} $-^A$ is the right adjoint to $-\oX A$. That is, the exponential to $A$ is characterised by an isomorphism
\[
\ladj{-} : \C (X\oX A, B) \cong \C (X,B^A) : \radj{-}
\]%
natural in $X$ and $B$.  We call the counit of the adjunction $\ev_B =
\radj{\id_{B^A}} : B^A\oX A \to B$ the \emph{evaluation
morphism} of the exponential.
%
When the exponential to $A$ exists, we say that $A$ is \emph{\exponentiable}.
When the exponential exists for every object we say that the monoidal category
\emph{has exponentials} or that it is a \emph{right-closed} monoidal category.
\end{mdefinition}

The next lemmata will be used in the proofs that follow.
\begin{mlemma}
\label{lemma:eval}
  Let $A, B, C$ be objects of a monoidal category $(\C,\oX,I,\alpha,\lambda,\rho)$,
  such that the exponential $-^A$ exists. For every $f : B \oX A \to C$, we have
  \[
    \ev \circ (\ladj{f} \oX \id) = f
  \]%
\end{mlemma}
\extended{
\begin{mproof}
\[
\begin{align*}
   & \ev \circ (\ladj{f} \oX \id) \\ 
=  &  \comment{definition $\ev$} \\
   & \radj{id} \circ (\ladj{f} \oX \id) \\
=  &  \comment{definition $(- \oX A)$ functor} \\
   & \radj{id} \circ (- \oX A)(\ladj{f}) \\
=  &  \comment{adjunction} \\
   & \radj{\ladj{f}} \\
=  &  \comment{adjunction} \\
   & f
\end{align*}
\]
\end{mproof}
}
\begin{mlemma}
\label{lemma:curry}
  Let $A, B, C, D$ be objects of a monoidal category $(\C,\oX,I,\alpha,\lambda,\rho)$,
  such that the exponential $-^C$ exists. For every $f : B \oX C \to D$ and $g : A \to B$, we have
  \[
    \ladj{f \circ (g \oX \id)} = \ladj{f} \circ g
  \]%
\end{mlemma}

\extended{
\hl{Innecesariamente larga, tal vez mejor usar naturalidad de $\ladj{-}$ directamente}
\begin{mproof}
\[
\begin{align*}
   & \ladj{f \circ (g \oX \id)} \\ 
=  &  \comment{definition $(- \oX C)$ functor} \\
   & \ladj{f \circ (- \oX C)(g)} \\
=  &  \comment{adjunction} \\
   & (f \circ (- \oX C)(g))^C \circ \ladj{\id} \\
=  &  \comment{$(-^C)$ functor} \\
   & f^C \circ ((- \oX C)(g))^C \circ \ladj{\id} \\
=  &  \comment{naturality} \\
   & f^C \circ \ladj{\id} \circ g \\
=  &  \comment{adjunction} \\
   & \ladj{f} \circ g \\
\end{align*}
\]
\end{mproof}
}


\subsection{Monoids in Monoidal Categories}

With the definition of monoidal category in place we may define a
monoid in such a category.

\begin{mdefinition}[Monoid]\label{def-mon-mon}
  A \emph{monoid} in a monoidal category
  $(\C,\oX,I,\alpha,\lambda,\rho)$ is a tuple $(M,m,e)$ where $M\in\C$
  and $m$ and $e$ are morphisms in $\C$
  $$\xymatrix{\oI\ar[r]^-e & M & \ar[l]_-m M\oX M}$$ 
such that the following diagrams commute.
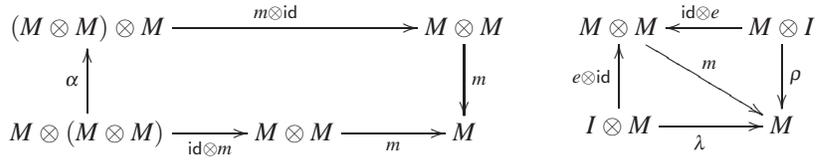
\[
\xymatrix@C+=10mm{
(M\oX M)\oX M\ar[rr]^{m\oX \id}& & M\oX M \ar[d]^{m}\\
M\oX (M\oX M) \ar[u]^{\alpha} \ar[r]_-{\id\oX m}
& M\oX M \ar[r]_-{m} & M \\
}\qquad
\xymatrix@C+=10mm{
    M \oX M \ar[dr]^{m} 
 &
   M\oX\oI \ar[l]_{\id\oX e} \ar[d]^{\rho}
\\
   \oI\oX M \ar[u]^{e\oX \id} \ar[r]_{\lambda}
 &
   M
}
\]

A \emph{monoid homomorphism} is an arrow
  $\xymatrix@1{M_1\ar[r]^f& M_2}$ in $\C$ such that the diagrams
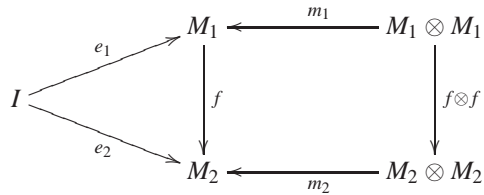
\[
\xymatrix@R=5mm@C+=20mm{
&M_1 \ar[dd]^{f}
&\ar[l]_{m_1} M_1\oX M_1 \ar[dd]^{f\oX f}
\\
\oI\ar[ru]^{e_1} \ar[rd]_{e_2}
\\
&M_2&\ar[l]^{m_2}M_2\oX M_2\\
}
\]
commute.
%
\end{mdefinition}

In the same manner that $A^*$ (the words on $A$) is the free monoid on
a set $A$, we can define the notion of free monoid in terms of
monoidal categories.

\begin{mdefinition}[Free Monoid]
Let $(\C,\oX,I,\alpha,\lambda,\rho)$ be a monoidal category.
The \emph{free monoid}
on an object $X$ in $\C$ is a monoid $(F, m_F, e_F)$ together with a morphism
$\ins : X \to F$ such that for any monoid $(G, m_G, e_G)$
and any morphism $f : X \to G$,
there exists a unique monoid homomorphism $\free{f} : F \to G$
that makes the following diagram commute.
\[
\xymatrix@R+=10mm@C+=15mm{
X \ar[r]^{\ins} \ar[rd]_{f}& F \ar@{-->}[d]^{\free{f}} \\
                               & G \\
}
\]
The morphism $\ins$ is called the \emph{insertion of generators} into
the free monoid.
\end{mdefinition}

Monoids in a monoidal category $\C$ and monoid homomorphisms form the
category $\Monoid{\C}$.
When the left-adjoint $(-)^*$ to the forgetful functor
$U:\Monoid{\C}\to\C$ exists, it maps an object $X$ to the free monoid
on $X$. There are several conditions that guarantee the existence of
free monoids~\cite{Dubuc74,Kelly80,Lack10}. Of particular importance
to us is the following:
\begin{mprop}
  \label{prop:free}
  Let $(\C,\oX,I,\alpha,\lambda,\rho)$ be a monoidal category with
  exponentials.  If $\C$ has binary coproducts, and for each $A \in
  \C$ the initial algebra for the endofunctor $\oI + A\oX −$ exists,
  then the monoid $A^*$ exists and its carrier is given by $\mu X.\
  \oI + A \oX X$.
\end{mprop}

\begin{mproof}
A multiplication on $A^*$ has the form $m : A^* \oX A^* \to A^*$.
By definition~\ref{def:exp}, it is equivalent to define a morphism
$A^* \to {A^*}^{A^*}$\!\!, and then use $\radj{-}$ to get $m$.
Exploiting the universal property of initial algebras, we define such
morphism by providing and algebra $I + A\oX {A^*}^{A^*} \to
{A^*}^{A^*}$\!\!, given by%
\footnote{For given $f : A \to C$ and $g : B \to C$, then $[f,g]$ is
  the unique morphism $A + B \to C$ such that $[f,g] \circ \inl = f$
  and $[f, g] \circ \inr = g$.}  $\left[ \ladj{\lambda_{A^*}},
  \ladj{\delta \circ \inr \circ (\id \oX \ev) \circ \alpha} \right]$
where $\delta : \oI + A\oX A^* \cong A^*$ is the initial algebra
structure over $A^*$.

The monoid structure on $A^*$ is then
\[
\begin{aligned}
e &= \delta \circ \inl \\
m &= \radj{\banana{[\ladj{\lambda_{A^*}}, \ladj{\delta \circ \inr \circ (\id \oX \ev) \circ \alpha}]}} \\
\end{aligned}
\]%
where the banana brackets $\banana{-}$ denote the universal morphism
from an initial algebra~\cite{meijer91}.

The insertion of generators and the universal morphism from the free
monoid to the monoid $(G, m_G, e_G)$ for $f : A \to G$ are:
\[
\begin{aligned}
\ins &= \delta \circ \inr \circ (\id \oX e) \circ \rho^{-1} \\
\free{f} &= \banana{[e_G,m_G \circ (f \oX \id)]}
\end{aligned}
\]%
\end{mproof}

\begin{coder}
  It is well known that the free monoid over a set $A$ is the set of
  lists of $A$. Unsurprisingly, when implementing in Haskell the
  formula of proposition~\ref{prop:free} for the case of $\Set$
  monoids, we obtain lists.
\begin{hscode}\SaveRestoreHook
\column{B}{@{}>{\hspre}l<{\hspost}@{}}%
\column{E}{@{}>{\hspre}l<{\hspost}@{}}%
\>[B]{}\mathkw{data}\;\mathsf{List}\;\Varid{a}\mathrel{=}\mathsf{Nil}\mid \mathsf{Cons}\;(\Varid{a},\mathsf{List}\;\Varid{a}){}\<[E]%
\ColumnHook
\end{hscode}\resethooks
\end{coder}


\begin{mdefinition}[Sub-monoid]
  Given a monoid $(M,e,m)$ in $\C$, and
a monic
  $i : M'\hookrightarrow M$ in $\C$, such that for some (unique) maps $e'$
  and $m'$, we have a commuting diagram
\[
\xymatrix@C+=15mm{
  \oI \ar[r]^{e}
  \ar@{=}[d]
&  M 
& \ar[l]_{m} M\oX M
\\
 \oI \ar[r]_{e'} 
 & M' \ar[u]_{i} 
 & M'\oX M' \ar[u]_{i\oX i}
    \ar[l]^{m'} 
}\]
then $(M',e',m')$ is a monoid, called the
\emph{sub-monoid} of $M$ induced by the monic $i$, and
$i$ is a monoid monomorphism from $M'$ to $M$.  
\end{mdefinition}


\subsection{Cayley Representation of a Monoid}

Every exponent in a monoidal category induces a monoid
of endomorphisms:
\begin{mdefinition}[Monoid of endomorphisms]
Let $(\C,\oX,I,\alpha,\lambda,\rho)$ be a monoidal category. The \emph{monoid
    of endomorphisms} on any 
  exponent $A\in\C$ is given by the
  diagram 
$$\xymatrix{
     \oI 
     \ar[r]^{i_A} 
   & \Rexp{A}{A}
   & \ar[l]_-{c_A}
     \Rexp{A}{A}\oX\Rexp{A}{A}
}
$$%
where
\begin{align*}
  i_A&=\ladj{\xymatrix{\oI\oX A\ar[r]^-{\lambda_A} & A}}\\
  c_A&=\ladj{\xymatrix@C+=15mm{
    (\Rexp{A}{A} \oX \Rexp{A}{A})\oX A
   \ar[r]^-{\alpha^{-1}} & \Rexp{A}{A}\oX(\Rexp{A}{A}\oX A) 
   \ar[r]^-{\id_{\Rexp{A}{A}} \oX\, \ev_A} & \Rexp{A}{A}\oX A \ar[r]^-{\ev_A} & A} }
\end{align*}
\end{mdefinition}
The Cayley representation theorem tell us that every monoid $(M,m,e)$ in a
monoidal category is a sub-monoid of a monoid of
endomorphisms whenever $M$ is {\exponentiable}.

\begin{mtheorem}[Cayley]\label{thm:Cayley}
  Let $(\C,\oX,I,\alpha,\lambda,\rho)$ be a monoidal category, and let
  $(M,e,m)$ be a monoid in $\C$. If $M$ is \exponentiable{} then
  $(M,e,m)$ is a sub-monoid of the monoid of endomorphisms
  $(\Rexp{M}{M},c_M,i_M)$, as witnessed by the monic $\rep = \ladj{m}
  : M \hookrightarrow \Rexp{M}{M}$. Moreover, $\abs\circ\rep =
  \id_{M}$ where $\abs$ is given by
\[\abs = \xymatrix@1@C+=15mm{\Rexp{M}{M} \ar[r]^-{\rho^{-1}_{\Rexp{M}{M}}} & 
                  \Rexp{M}{M}\oX\oI \ar[r]^{\id_{\Rexp{M}{M}}\oX\, e} &
                  \Rexp{M}{M}\oX M \ar[r]^-{\ev} &
                  M
                 }
\]%
%
\end{mtheorem}
\begin{mproof}
  The morphism $\rep: \xymatrix{ M \ar[r]^-{\ladj{m}} &
    \Rexp{M}{M}}$ is a monoid morphism.
\begin{align*}
   & \ladj{m} \circ e_M \\[\nl] 
=  &  \comment{lemma~\ref{lemma:curry}} \\[\nl]
   & \ladj{m \circ (e_M \oX \id)} \\[\nl]
=  &  \comment{monoid} \\[\nl]
   & \ladj{\lambda_M} \\[\nl]
=  &  \comment{definition of $i_M$} \\[\nl]
   & i_M
\end{align*}
\begin{align*}
   & c_M \circ (\ladj{m} \oX \ladj{m}) \\[\nl] 
=  &  \comment{definition $c_M$}\\[\nl]
   & \ladj{\ev \circ (\id_{M^M} \oX \ev) \circ \alpha^{-1}} \circ \ladj{m} \oX \ladj{m} \\[\nl]
=  &  \comment{lemma~\ref{lemma:curry}} \\[\nl]
   & \ladj{\ev \circ (\id_{M^M} \oX \ev) \circ \alpha^{-1} \circ ((\ladj{m} \oX \ladj{m}) \oX \id_M) } \\[\nl]
=  &  \comment{naturality $\alpha^{-1}$} \\[\nl]
   & \ladj{\ev \circ (\id_{M^M} \oX \ev) \circ (\ladj{m} \oX (\ladj{m} \oX \id_M)) \circ \alpha^{-1}} \\[\nl]
=  &  \comment{lemma~\ref{lemma:eval}} \\[\nl]
   & \ladj{\ev \circ (\ladj{m} \oX m) \circ \alpha^{-1}} \\[\nl]
=  &  \comment{lemma~\ref{lemma:eval}} \\[\nl]
   & \ladj{m \circ (\id_M \oX m) \circ \alpha^{-1}} \\[\nl]
=  &  \comment{monoid} \\[\nl]
   & \ladj{m \circ (m \oX \id_M)} \\[\nl]
=  &  \comment{lemma~\ref{lemma:curry}} \\[\nl]
   & \ladj{m} \circ m
\end{align*}

%
We have $\abs\circ\rep=\id_{M}$, and hence $\rep$ is monic.
\begin{align*}
   & \abs \circ \rep \\[\nl] 
=  &  \comment{definition of $\abs$ and $\rep$} \\[\nl]
   & \ev \circ (\id_{M^M} \oX e_M) \circ \rho^{-1}_{M^M} \circ \ladj{m} \\[\nl]
=  &  \comment{naturality of $\rho^{-1}$} \\[\nl]
   & \ev \circ (\id_{M^M} \oX e_M) \circ (\ladj{m} \oX \id) \circ \rho^{-1}_M \\[\nl]
=  &  \comment{tensor} \\[\nl]
   & \ev \circ (\ladj{m} \oX \id_M) \circ (\id_M \oX e_M) \circ \rho^{-1}_M \\[\nl]
=  &  \comment{lemma~\ref{lemma:eval}} \\[\nl]
   & m \circ (\id_M \oX e_M )\circ \rho^{-1}_M \\[\nl]
=  &  \comment{monoid} \\[\nl]
   & \rho_M \circ \rho^{-1}_M \\[\nl]
=  &  \comment{isomorphism} \\[\nl]
   & \id_M
\end{align*}
\end{mproof}

The Cayley theorem for sets (Theorem \ref{thm:Cayley_set}) is an instance of
this theorem for the category $\Set$. As new monoidal categories are introduced
in the following sections, more instances will be presented.




%

\section{Monads as Monoids}
\label{sec:monads}
Consider the (strict) monoidal category $\Endo = ([\Set,\Set],\circ,\Id)$ of
endofunctors, functor composition and identity functor. 
A monoid in this category consists of
\begin{itemize}
\item An endofunctor $M$, 
\item a natural transformation $m : M \circ M \to M$,
\item and a unit $e : \Id \to M$; such that the diagrams

\[
\xymatrix@C+=10mm{
(M\circ M)\circ M\ar[rr]^{m M}& & M\circ M \ar[d]^{m}\\
M\circ (M\circ M) \ar@{=}[u] \ar[r]_-{M m}
& M\circ M \ar[r]_-{m} & M \\
}\qquad
\xymatrix@C+=10mm{
    M \circ M \ar[dr]^{m} 
 &
   M\circ \Id \ar[l]_-{M e} \ar@{=}[d]
\\
   \Id\circ M \ar[u]^-{e M} \ar@{=}[r]
 &
   M
}
\]
commute.
\end{itemize}

Hence, a monoid in $\Endo$ is none other than a monad. Hence
the following often-heard slogan:
\emph{A monad is a monoid in a category of endofunctors}.

\begin{coder}
The corresponding implementation in Haskell is the following type class:
\begin{hscode}\SaveRestoreHook
\column{B}{@{}>{\hspre}l<{\hspost}@{}}%
\column{4}{@{}>{\hspre}l<{\hspost}@{}}%
\column{10}{@{}>{\hspre}l<{\hspost}@{}}%
\column{E}{@{}>{\hspre}l<{\hspost}@{}}%
\>[B]{}\mathkw{class}\;\mathsf{Functor}\;\Varid{m}\Rightarrow \mathsf{Triple}\;\Varid{m}\;\mathkw{where}{}\<[E]%
\\
\>[B]{}\hsindent{4}{}\<[4]%
\>[4]{}\eta {}\<[10]%
\>[10]{}\mathbin{::}\Varid{a}\to \Varid{m}\;\Varid{a}{}\<[E]%
\\
\>[B]{}\hsindent{4}{}\<[4]%
\>[4]{}\FN{join}{}\<[10]%
\>[10]{}\mathbin{::}\Varid{m}\;(\Varid{m}\;\Varid{a})\to \Varid{m}\;\Varid{a}{}\<[E]%
\ColumnHook
\end{hscode}\resethooks
\noindent where we have called the type class \ensuremath{\mathsf{Triple}} in order to not
clash with standard nomenclature which uses the name \ensuremath{\mathsf{Monad}} for the
presentation of a monad through its Kleisli extension:
\begin{hscode}\SaveRestoreHook
\column{B}{@{}>{\hspre}l<{\hspost}@{}}%
\column{4}{@{}>{\hspre}l<{\hspost}@{}}%
\column{12}{@{}>{\hspre}l<{\hspost}@{}}%
\column{E}{@{}>{\hspre}l<{\hspost}@{}}%
\>[B]{}\mathkw{class}\;\mathsf{Monad}\;\Varid{m}\;\mathkw{where}{}\<[E]%
\\
\>[B]{}\hsindent{4}{}\<[4]%
\>[4]{}\FN{return}{}\<[12]%
\>[12]{}\mathbin{::}\Varid{a}\to \Varid{m}\;\Varid{a}{}\<[E]%
\\
\>[B]{}\hsindent{4}{}\<[4]%
\>[4]{}(\bind ){}\<[12]%
\>[12]{}\mathbin{::}\Varid{m}\;\Varid{a}\to (\Varid{a}\to \Varid{m}\;\Varid{b})\to \Varid{m}\;\Varid{b}{}\<[E]%
\ColumnHook
\end{hscode}\resethooks

The latter has the advantage of not needing a \ensuremath{\mathsf{Functor}} instance and of
being easier to use when programming.
The two presentations are equivalent, as one can be obtained from the
other by taking \ensuremath{\eta \mathrel{=}\FN{return}}, \ensuremath{\FN{join}\mathrel{=}(\bind \FN{id})}, and \ensuremath{(\bind \Varid{f})\mathrel{=}\FN{join}\circ\FN{fmap}\;\Varid{f}}.
\end{coder}

\subsection{Exponential for Monads}
\label{sec:monad_exponential}

Finding an exponential in this category means finding a functor
$(-)^F$, such that we have an isomorphism natural in $G$ and $H$
\begin{equation}
  \label{eq:adj_functor_comp}
\Nat{H \circ F}{G} \quad \cong \quad \Nat{H}{G^F}  
\end{equation}

A useful technique for finding exponentials such as $G^F$ in a functor
category is to turn to the famous Yoneda lemma.

\begin{mtheorem}[Yoneda]
Let $\C$ be a locally small category. Then, there is an isomorphism
\[
   F\,X
  \quad\cong\quad 
  \Nat{\Hom_{\C}(X,-)}{F} 
\]%
natural in object $X : \C$ and functor $F: \C \to \Set$.  
That is, the set $F\,X$ is naturally isomorphic to the set of natural
transformations between the functor $\Hom_{\C}(X,-)$ and the
functor $F$.  
\end{mtheorem}

Now, if an exponential $G^F$ exists in the strict monoidal category
$([\Set,\Set],\circ,\Id)$, then the following must hold:
\begin{align*}
G^F\,X &\cong \Nat{\Hom(X, -)}{G^F} \\[\nl]
       &\cong \Nat{\Hom(X, -) \circ F}{G} 
\end{align*}
where the first isomorphism is by Yoneda, and the second is by
equation~\ref{eq:adj_functor_comp}. Therefore, whenever the expression
$\Nat{\Hom(X, -) \circ F}{G}$ makes sense, it can be taken to be the
\emph{definition} of the exponential $G^F$. Making sense in this case
means that the collection of natural transformations between $\Hom(X,
-) \circ F$ and $G$ is a set. The collection $\Nat{F}{G}$ of natural
transformation between two $\Set$ endofunctors $F$ and $G$ is not
always a set, i.e. $[\Set,\Set]$ is not locally small.
However, a sufficient condition for $\Nat{F}{G}$ to be a set is for
$F$ to be small. Small functors~\cite{Day2007651} are endofunctors on
$\Set$ which are a left Kan extension along the inclusion from a small
subcategory.  Therefore, every small functor $F$ is {\exponentiable}
in $\Endo$, with the exponential $(-)^F$ given by
\[
  G^F\,X = \Nat{\Hom(X, -) \circ F}{G} 
\]%

\begin{mremark}
  The functor $(-)^F$ is a right adjoint to the functor
  $(-\circ F)$ and is known as the right Kan extension along $F$.
\end{mremark}

\begin{coder}
The Haskell implementation of the exponential with respect to functor
composition is the following.

\begin{hscode}\SaveRestoreHook
\column{B}{@{}>{\hspre}l<{\hspost}@{}}%
\column{E}{@{}>{\hspre}l<{\hspost}@{}}%
\>[B]{}\mathkw{data}\;\mathsf{Exp}\;\Varid{f}\;\Varid{g}\;\Varid{x}\mathrel{=}\mathsf{Exp}\;(\forall \Varid{y}\hsforall .\ (\Varid{x}\to \Varid{f}\;\Varid{y})\to \Varid{g}\;\Varid{y}){}\<[E]%
\ColumnHook
\end{hscode}\resethooks

The components of isomorphism~\ref{eq:adj_functor_comp} are:
\begin{hscode}\SaveRestoreHook
\column{B}{@{}>{\hspre}l<{\hspost}@{}}%
\column{7}{@{}>{\hspre}l<{\hspost}@{}}%
\column{14}{@{}>{\hspre}l<{\hspost}@{}}%
\column{15}{@{}>{\hspre}l<{\hspost}@{}}%
\column{E}{@{}>{\hspre}l<{\hspost}@{}}%
\>[B]{}\varphi {}\<[15]%
\>[15]{}\mathbin{::}\mathsf{Functor}\;\Varid{h}\Rightarrow (\forall \Varid{x}\hsforall .\ \Varid{h}\;(\Varid{f}\;\Varid{x})\to \Varid{g}\;\Varid{x})\to \Varid{h}\;\Varid{y}\to \mathsf{Exp}\;\Varid{f}\;\Varid{g}\;\Varid{y}{}\<[E]%
\\
\>[B]{}\varphi \;{}\<[7]%
\>[7]{}\Varid{t}\;\Varid{y}{}\<[15]%
\>[15]{}\mathrel{=}\mathsf{Exp}\;(\lambda \Varid{k}\to \Varid{t}\;(\FN{fmap}\;\Varid{k}\;\Varid{y})){}\<[E]%
\\[\blanklineskip]%
\>[B]{}\varphi^{-1}{}\<[14]%
\>[14]{}\mathbin{::}(\forall \Varid{y}\hsforall .\ \Varid{h}\;\Varid{y}\to \mathsf{Exp}\;\Varid{f}\;\Varid{g}\;\Varid{y})\to \Varid{h}\;(\Varid{f}\;\Varid{x})\to \Varid{g}\;\Varid{x}{}\<[E]%
\\
\>[B]{}\varphi^{-1}\;{}\<[7]%
\>[7]{}\Varid{t}\;\Varid{x}{}\<[14]%
\>[14]{}\mathrel{=}\mathkw{let}\;\mathsf{Exp}\;\Varid{g}\mathrel{=}\Varid{t}\;\Varid{x}\;\mathkw{in}\;\Varid{g}\;\FN{id}{}\<[E]%
\ColumnHook
\end{hscode}\resethooks

\end{coder}

\subsection{Free Monads}

By restricting $\Endo$ to finitary functors we obtain the locally small, right-closed monoidal category $\Endo_F$~\cite{powerkelly}. In this category, we may apply
proposition~\ref{prop:free} and obtain the usual formula for the
free monad of an endofunctor $F$.
\[
F^*X  ~\cong~ X + F(F^* X)
\]%


\begin{coder}
  The formula above can be readily implemented by the
  datatype:
\begin{hscode}\SaveRestoreHook
\column{B}{@{}>{\hspre}l<{\hspost}@{}}%
\column{16}{@{}>{\hspre}c<{\hspost}@{}}%
\column{16E}{@{}l@{}}%
\column{19}{@{}>{\hspre}l<{\hspost}@{}}%
\column{E}{@{}>{\hspre}l<{\hspost}@{}}%
\>[B]{}\mathkw{data}\;\FN{Free}_\circ\;\Varid{f}\;\Varid{x}{}\<[16]%
\>[16]{}\mathrel{=}{}\<[16E]%
\>[19]{}\mathsf{Ret}\;\Varid{x}{}\<[E]%
\\
\>[16]{}\mid {}\<[16E]%
\>[19]{}\mathsf{Con}\;(\Varid{f}\;(\FN{Free}_\circ\;\Varid{f}\;\Varid{x})){}\<[E]%
\ColumnHook
\end{hscode}\resethooks
\noindent with monad instance:
\begin{hscode}\SaveRestoreHook
\column{B}{@{}>{\hspre}l<{\hspost}@{}}%
\column{3}{@{}>{\hspre}l<{\hspost}@{}}%
\column{12}{@{}>{\hspre}l<{\hspost}@{}}%
\column{19}{@{}>{\hspre}l<{\hspost}@{}}%
\column{E}{@{}>{\hspre}l<{\hspost}@{}}%
\>[B]{}\mathkw{instance}\;\mathsf{Functor}\;\Varid{f}\Rightarrow \mathsf{Monad}\;(\FN{Free}_\circ\;\Varid{f})\;\mathkw{where}{}\<[E]%
\\
\>[B]{}\hsindent{3}{}\<[3]%
\>[3]{}\FN{return}\;\Varid{x}{}\<[19]%
\>[19]{}\mathrel{=}\mathsf{Ret}\;\Varid{x}{}\<[E]%
\\
\>[B]{}\hsindent{3}{}\<[3]%
\>[3]{}(\mathsf{Ret}\;\Varid{x}){}\<[12]%
\>[12]{}\bind \Varid{f}{}\<[19]%
\>[19]{}\mathrel{=}\Varid{f}\;\Varid{x}{}\<[E]%
\\
\>[B]{}\hsindent{3}{}\<[3]%
\>[3]{}(\mathsf{Con}\;\Varid{m}){}\<[12]%
\>[12]{}\bind \Varid{f}{}\<[19]%
\>[19]{}\mathrel{=}\mathsf{Con}\;(\FN{fmap}\;(\bind \Varid{f})\;\Varid{m}){}\<[E]%
\ColumnHook
\end{hscode}\resethooks

There is no need to check that the instance satisfies the monad laws
since the definition is derived from Proposition~\ref{prop:free}.

The insertion of generators and the universal morphism from the free
monad are:
\begin{hscode}\SaveRestoreHook
\column{B}{@{}>{\hspre}l<{\hspost}@{}}%
\column{17}{@{}>{\hspre}l<{\hspost}@{}}%
\column{E}{@{}>{\hspre}l<{\hspost}@{}}%
\>[B]{}\ins\mathbin{::}\mathsf{Functor}\;\Varid{f}\Rightarrow \Varid{f}\xrightarrow{\sbt}\FN{Free}_\circ\;\Varid{f}{}\<[E]%
\\
\>[B]{}\ins\;\Varid{x}\mathrel{=}\mathsf{Con}\;(\FN{fmap}\;\mathsf{Ret}\;\Varid{x}){}\<[E]%
\\[\blanklineskip]%
\>[B]{}\FN{free}\mathbin{::}(\mathsf{Functor}\;\Varid{f},\mathsf{Monad}\;\Varid{m})\Rightarrow (\Varid{f}\circ\Varid{m}\xrightarrow{\sbt}\Varid{m})\to (\FN{Free}_\circ\;\Varid{f}\xrightarrow{\sbt}\Varid{m}){}\<[E]%
\\
\>[B]{}\FN{free}\;\Varid{f}\;(\mathsf{Ret}\;\Varid{x}){}\<[17]%
\>[17]{}\mathrel{=}\FN{return}\;\Varid{x}{}\<[E]%
\\
\>[B]{}\FN{free}\;\Varid{f}\;(\mathsf{Con}\;\Varid{t}){}\<[17]%
\>[17]{}\mathrel{=}\Varid{f}\;(\FN{fmap}\;(\FN{free}\;\Varid{k})\;\Varid{t}){}\<[E]%
\ColumnHook
\end{hscode}\resethooks
where the \ensuremath{\circ} in the type signature of \ensuremath{\FN{free}} is functor composition.
\end{coder}

\subsection{Cayley Representation of Monads}

For {\exponentiable} $F$, we may apply theorem~\ref{thm:Cayley} and
obtain the monad of endomorphisms $F^F$, the monad morphism
$\rep$, and the natural transformation $\abs$.
The monad $F^F$ corresponding to the monoid of endomorphisms on a
functor $F$ receives the name of \emph{codensity monad} on
$F$~\cite{macLaneS:catwm}.  
\begin{coder}
The codensity monad is implemented by the following datatype.
\begin{hscode}\SaveRestoreHook
\column{B}{@{}>{\hspre}l<{\hspost}@{}}%
\column{3}{@{}>{\hspre}l<{\hspost}@{}}%
\column{18}{@{}>{\hspre}l<{\hspost}@{}}%
\column{E}{@{}>{\hspre}l<{\hspost}@{}}%
\>[B]{}\mathkw{type}\;\mathsf{Rep}\;\Varid{f}\mathrel{=}\mathsf{Exp}\;\Varid{f}\;\Varid{f}{}\<[E]%
\\[\blanklineskip]%
\>[B]{}\mathkw{instance}\;\mathsf{Monad}\;(\mathsf{Rep}\;\Varid{f})\;\mathkw{where}{}\<[E]%
\\
\>[B]{}\hsindent{3}{}\<[3]%
\>[3]{}\FN{return}\;\Varid{x}{}\<[18]%
\>[18]{}\mathrel{=}\mathsf{Exp}\;(\lambda \Varid{h}\to \Varid{h}\;\Varid{x}){}\<[E]%
\\
\>[B]{}\hsindent{3}{}\<[3]%
\>[3]{}(\mathsf{Exp}\;\Varid{m})\bind \Varid{f}{}\<[18]%
\>[18]{}\mathrel{=}\mathsf{Exp}\;(\lambda \Varid{h}\to \Varid{m}\;(\lambda \Varid{x}\to \mathkw{let}\;\mathsf{Exp}\;\Varid{t}\mathrel{=}\Varid{f}\;\Varid{x}\;\mathkw{in}\;\Varid{t}\;\Varid{h})){}\<[E]%
\ColumnHook
\end{hscode}\resethooks
There is no need to check that the instance satisfies the monad
laws since the definition is derived directly from the general
definition of the monoid of endomorphisms.

The morphisms converting from a monad \ensuremath{\Varid{m}} to \ensuremath{\mathsf{Rep}\;\Varid{m}} and back are
the following.

\begin{hscode}\SaveRestoreHook
\column{B}{@{}>{\hspre}l<{\hspost}@{}}%
\column{18}{@{}>{\hspre}l<{\hspost}@{}}%
\column{20}{@{}>{\hspre}l<{\hspost}@{}}%
\column{E}{@{}>{\hspre}l<{\hspost}@{}}%
\>[B]{}\rep\mathbin{::}\mathsf{Monad}\;\Varid{m}\Rightarrow \Varid{m}\;\Varid{x}\to \mathsf{Rep}\;\Varid{m}\;\Varid{x}{}\<[E]%
\\
\>[B]{}\rep\;\Varid{m}{}\<[18]%
\>[18]{}\mathrel{=}\mathsf{Exp}\;(\Varid{m}\bind ){}\<[E]%
\\[\blanklineskip]%
\>[B]{}\abs\mathbin{::}\mathsf{Monad}\;\Varid{m}\Rightarrow \mathsf{Rep}\;\Varid{m}\;\Varid{x}\to \Varid{m}\;\Varid{x}{}\<[E]%
\\
\>[B]{}\abs\;(\mathsf{Rep}\;\Varid{m}){}\<[20]%
\>[20]{}\mathrel{=}\Varid{m}\;\FN{return}{}\<[E]%
\ColumnHook
\end{hscode}\resethooks
By Theorem~\ref{thm:Cayley}, we know that \ensuremath{\abs\circ\rep\mathrel{=}\FN{id}}, and that
\ensuremath{\abs} is a monad morphism. Hence, we may change the representation of
monadic computations on \ensuremath{\Varid{m}}, and perform computations on \ensuremath{\mathsf{Rep}\;\Varid{m}}. This change of representation is exactly the optimisation
introduced by~Voigl\"ander~\shortcite{Voigtlander08} and shown correct
by Hutton et al.~\shortcite{F5}.

Therefore, difference lists and the codensity transformation are both
instances of the same change of representation: the Cayley
representation.
\end{coder}

%

\section{Ends and Coends}
\label{sec:ends}
In this section we review the concept of a special type of limit
called \emph{end}, and its dual \emph{coend}. These concepts will be
used in the development of the next sections.

A limit for a functor $F : \C \to \D$ is universal cone to $F$, where
a cone is a natural transformation $\Delta_D\to F$ from the functor which is
constantly $D$, for a $D\in\D$, into the functor $F$.

When working with functors with mixed variance $F : \C^{\op} \times \C
\to \D$, rather than considering its limit, one is usually interested
in its end.
And end for a functor $F : \C^{\op} \times \C \to \D$ is a universal
\emph{wedge} to $F$, where a wedge is a \emph{dinatural}
transformation $\Delta_D\to F$ from the functor which is
constantly $D$ for a $D\in\D$, into the functor $F$.

We make this precise with the following definitions:
\begin{mdefinition}
  A \emph{dinatural transformation} $\alpha : F \to G$ between two
  functors $F, G : \C^{\op} \times \C \to \D$ is a family of morphisms
  of the form $\alpha_C : F(C,C) \to G(C,C)$, one morphism for each $C \in \C$, such
  that for every morphism $f : C \to C'$ the following diagram commutes.
  
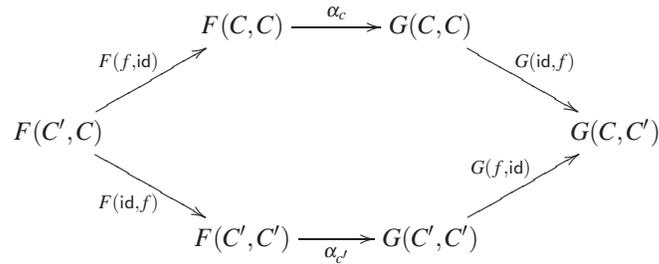
\[
  \xymatrix@C+=10mm{
    & F(C,C) \ar[r]^{\alpha_c}      & G(C,C) \ar[rd]^{G(\id,f)}   & \\
    F(C',C) \ar[ru]^{F(f,\id)} \ar[rd]_{F(\id,f)} &                            &                           & G(C,C') \\
    & F(C',C') \ar[r]_{\alpha_{c'}}  & G(C',C') \ar[ru]^{G(f,\id)}  &
  }
  \]%
\end{mdefinition}


An important difference between natural transformations and dinatural
transformations is that the latter can not be composed in the general
case.


\begin{mdefinition}
  A \emph{wedge} from an object $V \in \D$ to a functor $F : \C^{\op} \times
  \C \to \D$ is a dinatural transformation from the constant functor
  $\Delta_V : \C^{\op} \times \C \to \D$ to $F$. Explicitly, an object
  $V$ together with a family of morphisms $\alpha_X : V \to F(X,X)$
  such that for each $f : C \to C'$ the following diagram commutes.
  
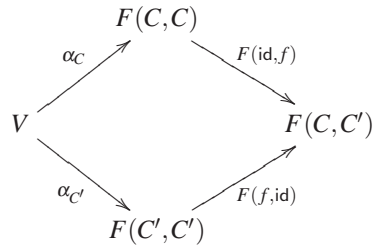
\[
  \xymatrix{
                                  & F(C,C) \ar[dr]^{F(\id,f)}    & \\
    V \ar[ur]^{\alpha_C} \ar[dr]_{\alpha_{C'}}  &                             & F(C,C') \\
                                  & F(C',C') \ar[ur]_{F(f,\id)}  &
  }
  \]

\end{mdefinition}

In the same way a limit is a final cone, we define an \emph{end} as a
final wedge.

\begin{mdefinition}
  The \emph{end} of a functor $F : \C^{\op} \times \C \to \D$ is a
  final wedge for $F$.  Explicitly, it is an object $V\in\D$ together
  with a family of morphisms $\omega_C : V \to F(C,C)$ such that the
  diagram
  
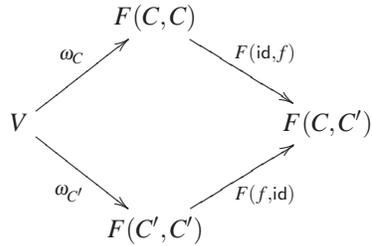
\[
  \xymatrix{
                                  & F(C,C) \ar[dr]^{F(\id,f)}    & \\
    V \ar[ur]^{\omega_C} \ar[dr]_{\omega_{C'}}  &                             & F(C,C') \\
                                  & F(C',C') \ar[ur]_{F(f,\id)}  &
  }
  \]%
  commutes for each $f : C \to C'$, and such that for every wedge from
  $V'\in\D$, given by a family of morphisms $\gamma_c : V' \to F(C,C)$
  such that $F(\id,f) \circ \gamma_c = F(f,\id) \circ \gamma_c'$ for
  every $f : C \to C'$, there exists a unique morphism $! : V' \to V$
  such that the following diagram commutes.
  
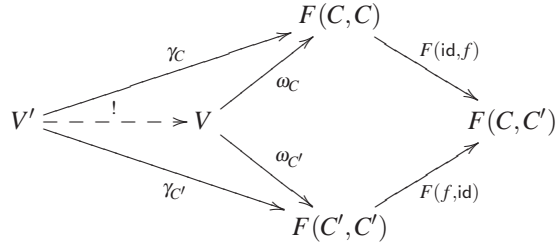
\[
  \xymatrix{
                                               & &                                  & F(C,C) \ar[dr]^{F(\id,f)}    & \\
    V' \ar[urrr]^{\gamma_C} \ar[drrr]_{\gamma_{C'}} \ar@{-->}[rr]^{!} & & V \ar[ur]_{\omega_C} \ar[dr]^{\omega_{C'}}  &                             & F(C,C') \\
                                               & &                                  & F(C',C') \ar[ur]_{F(f,\id)}  &
  }
  \]
\end{mdefinition}

The object $V$ is usually denoted by $\int_A F(A,A)$ and referred to
as ``the end of $F$''.

    

\begin{coder}
  One nice feature of ends is that it leads to a natural
  implementation of categorical concepts in Haskell by replacing the
  end by a universal quantifier. For example, the natural
  transformations between two functors $F$ and $G$ can be expressed as
  an end
\[
 \int_X FX \to GX
\]
(where by $X \to Y$ we note the exponential on $\Set$) and implemented as follows.
\begin{hscode}\SaveRestoreHook
\column{B}{@{}>{\hspre}l<{\hspost}@{}}%
\column{E}{@{}>{\hspre}l<{\hspost}@{}}%
\>[B]{}\mathkw{type}\;\Varid{f}\xrightarrow{\sbt}\Varid{g}\mathrel{=}\forall \Varid{x}\hsforall .\ \Varid{f}\;\Varid{x}\to \Varid{g}\;\Varid{x}{}\<[E]%
\ColumnHook
\end{hscode}\resethooks
\vspace{-6mm}
\end{coder}

Ends can be seen as a generalised product, but cut down by a relation
of dinaturality.  Following this view, a morphism to an end is defined
by a dinatural family of morphisms:

\[ \infer{\phi_X : Y \to F(X,X), \mbox{ dinatural in }X}{\langle \phi \rangle : Y \to \int_A F(A,A)} \]

\begin{mprop}
  By the universal property of ends, $\langle \phi \rangle$ is the
  unique morphism such that $\omega_X \circ \langle \phi \rangle =
  \phi_X$.
\end{mprop}

Given a dinatural transformation $\alpha : \Delta_Y \to F$, and a
morphism $h : Z \to Y$, the family of morphisms defined by $(\alpha
\circ h)_C = \alpha_C \circ h$ is dinatural in $C$. Using the
universal property of ends, we obtain the following proposition:

\begin{mprop}
  Let $\phi_X : Y \to F(X,X)$ be a  family of morphisms dinatural in $X$, and
  a let $h : Z \to Y$. Then $\langle \phi \circ h \rangle = \langle \phi \rangle \circ h$.
\end{mprop}

When defining a family of morphisms, abstracting over the varying
object comes in handy. We will use $\Lambda$ as a binder
for objects variables. For example, $(\alpha \circ h)_C = \alpha_C
\circ h$ can be defined directly as $\alpha \circ h = \Lambda C
.\, \alpha_C \circ h$,

There are dual notions of wedges and ends, namely cowedges and
coends. We briefly summarise their definitions.

\begin{mdefinition}
  A \emph{cowedge} from $F$ is an object $V$ together with a dinatural
  transformation $\alpha : F \to \Delta_V$.
\end{mdefinition}

\begin{mdefinition}
  A \emph{coend} is an initial cowedge. Explicitly, a coend of $F$ is
  an object $V$ together with a family of morphisms $\iota_C : F(C,C)
  \to V$ such that $\iota_X \circ F(f,\id) = \iota_Y \circ F(\id,f)$,
  which is universal with respect to this property: for every 
  object $V'$ and family of morphisms $\gamma_C : F(C,C) \to V'$ such
  that $\gamma_X \circ F(f,\id) = \gamma_Y \circ F(\id,f)$, then there
  exists a unique morphism $f : V \to V'$ such that $\gamma_X = f \circ
  \iota_X$.
\end{mdefinition}

A coend can be seen as a generalised coproduct, quotiented by an
equivalence relation.  If $\phi_X : F(X,X) \to Y$ is a family of
morphisms dinatural in $X$, then the morphism from $\int^A F(A,A)$ to
$Y$ given by the universal property of coends is denoted as $[ \phi
]$:
\[ \infer{\phi_X : F(X,X) \to Y, \mbox{ dinatural in }X}{[ \phi ] : \int_A F(A,A) \to Y} \]%

\begin{coder}
  In the same way an end can be implemented as a universal quantifier,
  a coend can be implemented as an existential quantifier, as
  supported by modern implementations of Haskell.
\end{coder}

We finish this section by presenting the Yoneda lemma in the language of
ends and coends. Focusing on functors $\C^{\op} \times \C\to\Set$,
with $\C$ a small category, we can form the set of dinatural transformations
between two such functors. The fact that such dinatural transformations
form a set is justified by the next proposition.

\begin{mprop}
  Let $F,G : \C^{\op} \times \C\to\Set$, with $\C$ a small
  category. Dinatural transformations from $F$ to $G$ are in a
  one-to-one correspondence with global elements of $\int_A F(A,A) \to
  G(A,A)$. If we denote the dinatural transformations
  between $F$ and $G$ by $\Dinat{F}{G}$, we obtain:
 \[ \Dinat{F}{G} \quad \cong \quad \int_A F(A,A) \to G(A,A) \]

 In particular, when $F$ and $G$ are functors in one covariant
 variable (i.e. dummy in their contravariant variable), $\Dinat{F}{G}$
 reduces to $\Nat{F}{G}$ and we have

  \[ \Nat{F}{G} \quad \cong \quad \int_A F(A) \to G(A) \]
\end{mprop}

The Yoneda lemma in its end and coend form~\cite{AsadaH10,Asada10} is
usually expressed as:
\[ F X \quad \cong \quad \int_Y \Hom_{\C}(X,Y)\to FY 
       \quad \cong \quad \int^Y F Y \times \Hom_{\C}(Y,X) \]%

\begin{coder}
  We can interpret the end form of Yoneda lemma as an isomorphism
  between types \ensuremath{\Varid{f}\;\Varid{x}} and \ensuremath{\forall \Varid{y}\hsforall .\ (\Varid{x}\to \Varid{y})\to \Varid{f}\;\Varid{y}} whenever \ensuremath{\Varid{f}} is
  a \ensuremath{\mathsf{Functor}}.

The components of the isomorphism are implemented as
\begin{hscode}\SaveRestoreHook
\column{B}{@{}>{\hspre}l<{\hspost}@{}}%
\column{8}{@{}>{\hspre}c<{\hspost}@{}}%
\column{8E}{@{}l@{}}%
\column{9}{@{}>{\hspre}c<{\hspost}@{}}%
\column{9E}{@{}l@{}}%
\column{12}{@{}>{\hspre}l<{\hspost}@{}}%
\column{13}{@{}>{\hspre}l<{\hspost}@{}}%
\column{E}{@{}>{\hspre}l<{\hspost}@{}}%
\>[B]{}\varphi {}\<[8]%
\>[8]{}\mathbin{::}{}\<[8E]%
\>[12]{}\mathsf{Functor}\;\Varid{f}\Rightarrow \Varid{f}\;\Varid{x}\to (\forall \Varid{y}\hsforall .\ (\Varid{x}\to \Varid{y})\to \Varid{f}\;\Varid{y}){}\<[E]%
\\
\>[B]{}\varphi \;\Varid{v}{}\<[8]%
\>[8]{}\mathrel{=}{}\<[8E]%
\>[12]{}\lambda \Varid{f}\to \FN{fmap}\;\Varid{f}\;\Varid{v}{}\<[E]%
\\[\blanklineskip]%
\>[B]{}\varphi^{-1}{}\<[9]%
\>[9]{}\mathbin{::}{}\<[9E]%
\>[13]{}(\forall \Varid{y}\hsforall .\ (\Varid{x}\to \Varid{y})\to \Varid{f}\;\Varid{y})\to \Varid{f}\;\Varid{x}{}\<[E]%
\\
\>[B]{}\varphi^{-1}\;\Varid{g}{}\<[9]%
\>[9]{}\mathrel{=}{}\<[9E]%
\>[13]{}\Varid{g}\;\FN{id}{}\<[E]%
\ColumnHook
\end{hscode}\resethooks

Similarly, its coend form (also known as ``coYoneda lemma'') is
expressed by

\begin{hscode}\SaveRestoreHook
\column{B}{@{}>{\hspre}l<{\hspost}@{}}%
\column{14}{@{}>{\hspre}l<{\hspost}@{}}%
\column{E}{@{}>{\hspre}l<{\hspost}@{}}%
\>[B]{}\psi {}\<[14]%
\>[14]{}\mathbin{::}\mathsf{Functor}\;\Varid{f}\Rightarrow \Varid{f}\;\Varid{x}\to (\exists\;\Varid{y}.\ (\Varid{f}\;\Varid{y},\Varid{y}\to \Varid{x})){}\<[E]%
\\
\>[B]{}\psi \;\Varid{v}{}\<[14]%
\>[14]{}\mathrel{=}(\Varid{v},\FN{id}){}\<[E]%
\\[\blanklineskip]%
\>[B]{}\psi^{-1}{}\<[14]%
\>[14]{}\mathbin{::}\mathsf{Functor}\;\Varid{f}\Rightarrow (\exists\;\Varid{y}.\ (\Varid{f}\;\Varid{y},\Varid{y}\to \Varid{x}))\to \Varid{f}\;\Varid{x}{}\<[E]%
\\
\>[B]{}\psi^{-1}\;(\Varid{x},\Varid{g}){}\<[14]%
\>[14]{}\mathrel{=}\FN{fmap}\;\Varid{g}\;\Varid{x}{}\<[E]%
\ColumnHook
\end{hscode}\resethooks
\end{coder}

%

\section{Applicatives as Monoids}
\label{sec:applicative}

Similarly to monads, applicative
functors~\cite{mcbride08:applicative-programming} are a class of
functors used to write certain effectful computations. These functors
come with an operation that allows evaluation of functions inside the
functor. Compared to monads, applicative functors are a strictly
weaker notion: every monad is an applicative functor (see
Section~\ref{sec:identity}), but there are applicative functors which
are not monads. The main difference between monads and applicative
functors is that the latter does not allow effects to depend on
previous values, i.e. they are fixed beforehand.

\begin{coder}
In Haskell, these functors are represented by the
following type class:

\begin{hscode}\SaveRestoreHook
\column{B}{@{}>{\hspre}l<{\hspost}@{}}%
\column{3}{@{}>{\hspre}l<{\hspost}@{}}%
\column{12}{@{}>{\hspre}l<{\hspost}@{}}%
\column{E}{@{}>{\hspre}l<{\hspost}@{}}%
\>[B]{}\mathkw{class}\;\mathsf{Functor}\;\Varid{f}\Rightarrow \mathsf{Applicative}\;\Varid{f}\;\mathkw{where}{}\<[E]%
\\
\>[B]{}\hsindent{3}{}\<[3]%
\>[3]{}\FN{pure}{}\<[12]%
\>[12]{}\mathbin{::}\Varid{a}\to \Varid{f}\;\Varid{a}{}\<[E]%
\\
\>[B]{}\hsindent{3}{}\<[3]%
\>[3]{}(\circledast){}\<[12]%
\>[12]{}\mathbin{::}\Varid{f}\;(\Varid{a}\to \Varid{b})\to \Varid{f}\;\Varid{a}\to \Varid{f}\;\Varid{b}{}\<[E]%
\ColumnHook
\end{hscode}\resethooks
\end{coder}

Since their introduction, applicative functors have been
characterised categorically as \emph{strong lax monoidal
  functors}~\cite{mcbride08:applicative-programming}.  We explain the
notions of \emph{strong functor} and \emph{lax monoidal functor}
separately.  In simple words, a lax monoidal functor is a functor
preserving the monoidal structure of the categories involved.

\begin{mdefinition}
  A \emph{lax monoidal functor} $F : \C \to \D$ is a functor between
  the underlying categories of two monoidal categories $(\C, \oX, \oI,
  \alpha_{\C}, \lambda_{\C}, \rho_{\C})$ and $(\D, \oP, \oJ,
  \alpha_{\D}, \lambda_{\D}, \rho_{\D})$ together with a natural
  transformation
\[ \phi_{A,B} : F(A) \oP F(B) \to F(A \oX B) \]%
and a morphism
\[ \eta : \oJ \to F(\oI) \]%
such that the following diagrams commute.
\[
\xymatrix@C+=10mm{
F A \oP (F B \oP F C) \ar[r]_{\id \oP \phi_{B,C}} \ar[d]_{\alpha_{\D}} & F A \oP F (B \oX C) \ar[r]_{\phi_{A, (B \oX C)}} & F (A \oX (B \oX C)) \ar[d]^{F \alpha_{\C}} \\
(F A \oP F B) \oP F C \ar[r]^{\phi_{A,B} \oP \id} & F (A \oX B) \oP F C \ar[r]^{\phi_{(A \oX B),C}} & F ((A \oX B) \oX C) \\
}
\]
\[
\xymatrix@C+=10mm{
F A \oP \oJ \ar[r]^{\id \oP \eta} \ar[d]_{\rho_{\D}} & F A \oP F \oI \ar[d]^{\phi_{A,\oI}} \\
F A & F (A \oX \oI) \ar[l]^{F \rho_{\C}}
}
\qquad
\xymatrix@C+=10mm{
\oJ \oP F A  \ar[r]^{\eta \oP \id} \ar[d]_{\lambda_{\D}} & F \oI \oP F A \ar[d]^{\phi_{\oI,A}} \\
F A & F (\oI \oX A) \ar[l]^{F \rho_{\C}}
}
\]
A \emph{monoidal functor} is a lax monoidal functor in which $\phi$ and $\eta$ are isomorphisms.
\end{mdefinition}

\begin{mdefinition}
An endofunctor $F : \C \to \C$ is \emph{strong} when it comes equipped with a natural transformation
\[ \funst_{A,B} : A \oX F B \to F (A \oX B) \]%
called a \emph{strength} such that following diagrams commute.
\[
\xymatrix@C+=8mm{
1 \oX F(A) \ar[d]_{\funst} \ar[dr]^{\rho} & \\
F(1 \oX A) \ar[r]_{F \rho} & F(A)
} \quad
\xymatrix@C+=8mm{
A \oX (B \oX F C) \ar[r]^{A \oX \funst} \ar[d]^{\alpha} & A \oX F (B \oX C) \ar[r]^{\funst} &  F (A \oX (B \oX C)) \ar[d]^{F(\alpha)} \\
(A \oX B) \oX F C  \ar[rr]_{\funst} & & F ((A \oX B) \oX C)
}
\]
\end{mdefinition}

All endofunctors on the (cartesian) monoidal category $\Set$ come with
a unique strength, so all functors in $[\Set, \Set]$ are strong.
Now, a \emph{strong lax monoidal functor} is simply a lax monoidal
functor which is also a strong functor and in which the strength
interacts coherently with the monoidal structure. In our setting of
$\Set$ endofunctors we get this coherence for free. 

\begin{coder}
  The categorical characterisation of applicative functors as strong
  lax monoidal functors gives rise to an alternative (but equivalent)
  implementation of applicative functors:
\begin{hscode}\SaveRestoreHook
\column{B}{@{}>{\hspre}l<{\hspost}@{}}%
\column{3}{@{}>{\hspre}l<{\hspost}@{}}%
\column{11}{@{}>{\hspre}l<{\hspost}@{}}%
\column{E}{@{}>{\hspre}l<{\hspost}@{}}%
\>[B]{}\mathkw{class}\;\mathsf{Functor}\;\Varid{f}\Rightarrow \mathsf{Monoidal}\;\Varid{f}\;\mathkw{where}{}\<[E]%
\\
\>[B]{}\hsindent{3}{}\<[3]%
\>[3]{}\FN{unit}{}\<[11]%
\>[11]{}\mathbin{::}\Varid{f}\;(){}\<[E]%
\\
\>[B]{}\hsindent{3}{}\<[3]%
\>[3]{}(\star){}\<[11]%
\>[11]{}\mathbin{::}\Varid{f}\;\Varid{a}\to \Varid{f}\;\Varid{b}\to \Varid{f}\;(\Varid{a},\Varid{b}){}\<[E]%
\ColumnHook
\end{hscode}\resethooks
\vspace{-5mm}
\end{coder}


We saw how monads are monoids in a particular monoidal category.
Applicative functors can be shown to be monoids too.  Interestingly,
they are monoids in the same category as monads: \emph{An applicative
  functor is a monoid in a category of endofunctors}.
However, it is not the same monoidal category, as this time we must
consider a different notion of tensor. For monads we used composition;
for applicative functors we use a tensor called \emph{Day
  convolution}~\cite{Day70}. Given a cartesian closed category $\C$,
two functors $F, G : \C \to \C$, and an object $B$ in $\C$, the Day
convolution $(F \dayt G) B$ is a new object in $\C$ defined as:

\[ (F \dayt G) B = \int^{C,D} F C \times G D \times B^{(C \times D)} \]


The coend does not necessarily exist for arbitrary $\Set$
endofunctors, but it is guaranteed to exist for small
functors~\cite{Day2007651}. Unless otherwise stated, in the remainder
of the section we will work with $[\Set, \Set]_S$ the category of
small $\Set$ endofunctors.

Applying theorem IX.7.1 of Mac Lane~\shortcite{macLaneS:catwm},
it can be shown that $F \dayt
G$ is not only a mapping between objects, but also a mapping between
morphisms, and that it respects the functor laws.
Furthermore, given natural transformations $\alpha : F \to G$ and
$\beta : H \to I$, we can form a natural transformation $\alpha \dayt
\beta : F \dayt H \to G \dayt I$.  This makes the Day convolution a
bifunctor $- \dayt - : [\Set, \Set]_S \times [\Set, \Set]_S \to [\Set,
\Set]_S$.

\begin{coder}

  The coend in the definition of the Day convolution can be
  implemented by an existential datatype. In the definition below,
  done in GADT style, the type variables \ensuremath{\Varid{c}} and \ensuremath{\Varid{d}} are existentially
  quantified.

\begin{hscode}\SaveRestoreHook
\column{B}{@{}>{\hspre}l<{\hspost}@{}}%
\column{3}{@{}>{\hspre}l<{\hspost}@{}}%
\column{E}{@{}>{\hspre}l<{\hspost}@{}}%
\>[B]{}\mathkw{data}\;(\Varid{f}\dayt\Varid{g})\;\Varid{b}\;\mathkw{where}{}\<[E]%
\\
\>[B]{}\hsindent{3}{}\<[3]%
\>[3]{}\mathsf{Day}\mathbin{::}\Varid{f}\;\Varid{c}\to \Varid{g}\;\Varid{d}\to ((\Varid{c},\Varid{d})\to \Varid{b})\to (\Varid{f}\dayt\Varid{g})\;\Varid{b}{}\<[E]%
\\[\blanklineskip]%
\>[B]{}\mathkw{instance}\;(\mathsf{Functor}\;\Varid{f},\mathsf{Functor}\;\Varid{g})\Rightarrow \mathsf{Functor}\;(\Varid{f}\dayt\Varid{g})\;\mathkw{where}{}\<[E]%
\\
\>[B]{}\hsindent{3}{}\<[3]%
\>[3]{}\FN{fmap}\;\Varid{f}\;(\mathsf{Day}\;\Varid{x}\;\Varid{y}\;\Varid{g})\mathrel{=}\mathsf{Day}\;\Varid{x}\;\Varid{y}\;(\Varid{f}\circ\Varid{g}){}\<[E]%
\ColumnHook
\end{hscode}\resethooks

The Day convolution is a bifunctor with the following mapping of morphisms:
\begin{hscode}\SaveRestoreHook
\column{B}{@{}>{\hspre}l<{\hspost}@{}}%
\column{E}{@{}>{\hspre}l<{\hspost}@{}}%
\>[B]{}\FN{bimap}\mathbin{::}(\Varid{f}\xrightarrow{\sbt}\Varid{h})\to (\Varid{g}\xrightarrow{\sbt}\Varid{i})\to (\Varid{f}\dayt\Varid{g}\xrightarrow{\sbt}\Varid{h}\dayt\Varid{i}){}\<[E]%
\\
\>[B]{}\FN{bimap}\;\Varid{m}_{1}\;\Varid{m}_{2}\;(\mathsf{Day}\;\Varid{x}\;\Varid{y}\;\Varid{f})\mathrel{=}\mathsf{Day}\;(\Varid{m}_{1}\;\Varid{x})\;(\Varid{m}_{2}\;\Varid{y})\;\Varid{f}{}\<[E]%
\ColumnHook
\end{hscode}\resethooks
\vspace{-5mm}
\end{coder}

The following proposition allows us to write morphisms from the image
of the Day convolution to another object.

\begin{mprop}
\label{prop:bij}
There is a one-to-one correspondence defining morphisms going out of
a Day convolution
\begin{equation}
  \label{eq:adj_day}
  [\C, \C](F \dayt G, H) \quad \stackrel{\vartheta}{\cong} \quad [\C \times \C, \C](\times \circ (F \times G), H \circ \times)
\end{equation}
which is natural in $F$, $G$, and $H$. Here, $\times : \C \times \C
\to \C$ is the functor which takes an object $(A,B)$ of the product
category into a product of objects $A \times B$.
\end{mprop}

\begin{mremark}[Day convolution as a left Kan extension]
  In view of the last proposition, the Day convolution $F \dayt G$ is the
  left Kan extension of $\times \circ (F \times
  G)$ along $\times$.
\end{mremark}

\begin{coder}
  The proposition above shows an equivalence between the types \ensuremath{(\Varid{f}\dayt\Varid{g})\xrightarrow{\sbt}\Varid{h}} and \ensuremath{\forall \Varid{a}\hsforall \;\Varid{b}.\ (\Varid{f}\;\Varid{a},\Varid{g}\;\Varid{b})\to \Varid{h}\;(\Varid{a},\Varid{b})}.
\begin{hscode}\SaveRestoreHook
\column{B}{@{}>{\hspre}l<{\hspost}@{}}%
\column{E}{@{}>{\hspre}l<{\hspost}@{}}%
\>[B]{}\vartheta \mathbin{::}(\Varid{f}\dayt\Varid{g}\xrightarrow{\sbt}\Varid{h})\to (\Varid{f}\;\Varid{a},\Varid{g}\;\Varid{b})\to \Varid{h}\;(\Varid{a},\Varid{b}){}\<[E]%
\\
\>[B]{}\vartheta \;\Varid{f}\;(\Varid{x},\Varid{y})\mathrel{=}\Varid{f}\;(\mathsf{Day}\;\Varid{x}\;\Varid{y}\;\FN{id}){}\<[E]%
\\[\blanklineskip]%
\>[B]{}\vartheta^{-1}\mathbin{::}\mathsf{Functor}\;\Varid{h}\Rightarrow (\forall \Varid{a}\hsforall \;\Varid{b}.\ (\Varid{f}\;\Varid{a},\Varid{g}\;\Varid{b})\to \Varid{h}\;(\Varid{a},\Varid{b}))\to (\Varid{f}\dayt\Varid{g}\xrightarrow{\sbt}\Varid{h}){}\<[E]%
\\
\>[B]{}\vartheta^{-1}\;\Varid{g}\;(\mathsf{Day}\;\Varid{x}\;\Varid{y}\;\Varid{f})\mathrel{=}\FN{fmap}\;\Varid{f}\;(\Varid{g}\;(\Varid{x},\Varid{y})){}\<[E]%
\ColumnHook
\end{hscode}\resethooks
\end{coder}

In contrast to the composition tensor, the Day convolution
is not strict. Moreover, the Day convolution is symmetric, which
together with appropriate natural transformations $\alpha$, $\lambda$
and $\rho$ make $\EndD = \left([\Set, \Set]_S, \dayt, \Id, \alpha,
  \lambda, \rho, \gamma\right)$ a symmetric monoidal
category~\cite{Day70}.

\begin{coder}
  
  Here we present the natural transformations of the monoidal category $\EndD$.
  In order to do that we first implement the identity functor.

\begin{hscode}\SaveRestoreHook
\column{B}{@{}>{\hspre}l<{\hspost}@{}}%
\column{12}{@{}>{\hspre}l<{\hspost}@{}}%
\column{18}{@{}>{\hspre}l<{\hspost}@{}}%
\column{21}{@{}>{\hspre}l<{\hspost}@{}}%
\column{28}{@{}>{\hspre}l<{\hspost}@{}}%
\column{E}{@{}>{\hspre}l<{\hspost}@{}}%
\>[B]{}\mathkw{data}\;\mathsf{Id}\;\Varid{a}\mathrel{=}\mathsf{Id}\;\Varid{a}\;\mathkw{deriving}\;\mathsf{Functor}{}\<[E]%
\\[\blanklineskip]%
\>[B]{}\lambda \mathbin{::}\mathsf{Functor}\;\Varid{f}\Rightarrow \Varid{f}\xrightarrow{\sbt}\mathsf{Id}\dayt\Varid{f}{}\<[E]%
\\
\>[B]{}\lambda \;\Varid{x}{}\<[12]%
\>[12]{}\mathrel{=}\mathsf{Day}\;(\mathsf{Id}\;())\;\Varid{x}\;\FN{snd}{}\<[E]%
\\[\blanklineskip]%
\>[B]{}\rho \mathbin{::}\mathsf{Functor}\;\Varid{f}\Rightarrow \Varid{f}\xrightarrow{\sbt}\Varid{f}\dayt\mathsf{Id}{}\<[E]%
\\
\>[B]{}\rho \;\Varid{x}{}\<[12]%
\>[12]{}\mathrel{=}\mathsf{Day}\;\Varid{x}\;(\mathsf{Id}\;())\;\FN{fst}{}\<[E]%
\\[\blanklineskip]%
\>[B]{}\alpha \mathbin{::}(\Varid{f}\dayt\Varid{g})\dayt\Varid{h}\xrightarrow{\sbt}\Varid{f}\dayt(\Varid{g}\dayt\Varid{h}){}\<[E]%
\\
\>[B]{}\alpha \;(\mathsf{Day}\;(\mathsf{Day}\;\Varid{x}\;\Varid{y}\;\Varid{f})\;\Varid{z}\;\Varid{g})\mathrel{=}\mathsf{Day}\;\Varid{x}\;(\mathsf{Day}\;\Varid{y}\;\Varid{z}\;\Varid{f}_{1})\;\Varid{f}_{2}{}\<[E]%
\\
\>[B]{}\hsindent{21}{}\<[21]%
\>[21]{}\mathkw{where}\;{}\<[28]%
\>[28]{}\Varid{f}_{1}\mathrel{=}\lambda (\Varid{d},\Varid{b})\to ((\lambda \Varid{c}\to \Varid{f}\;(\Varid{c},\Varid{d})),\Varid{b}){}\<[E]%
\\
\>[28]{}\Varid{f}_{2}\mathrel{=}\lambda (\Varid{c},(\Varid{h},\Varid{b}))\to \Varid{g}\;(\Varid{h}\;\Varid{c},\Varid{b}){}\<[E]%
\\[\blanklineskip]%
\>[B]{}\gamma \mathbin{::}(\Varid{f}\dayt\Varid{g})\xrightarrow{\sbt}(\Varid{g}\dayt\Varid{f}){}\<[E]%
\\
\>[B]{}\gamma \;(\mathsf{Day}\;\Varid{x}\;\Varid{y}\;\Varid{f})\mathrel{=}\mathsf{Day}\;\Varid{y}\;\Varid{x}\;(\Varid{f}\circ\FN{swap}){}\<[E]%
\\
\>[B]{}\hsindent{18}{}\<[18]%
\>[18]{}\mathkw{where}\;\FN{swap}\;(\Varid{x},\Varid{y})\mathrel{=}(\Varid{y},\Varid{x}){}\<[E]%
\ColumnHook
\end{hscode}\resethooks
\extended{ Their respective inverses are defined as:
\begin{hscode}\SaveRestoreHook
\column{B}{@{}>{\hspre}l<{\hspost}@{}}%
\column{21}{@{}>{\hspre}l<{\hspost}@{}}%
\column{28}{@{}>{\hspre}l<{\hspost}@{}}%
\column{32}{@{}>{\hspre}l<{\hspost}@{}}%
\column{E}{@{}>{\hspre}l<{\hspost}@{}}%
\>[B]{}\lambda^{-1}\mathbin{::}\mathsf{Functor}\;\Varid{f}\Rightarrow \mathsf{Id}\dayt\Varid{f}\xrightarrow{\sbt}\Varid{f}{}\<[E]%
\\
\>[B]{}\lambda^{-1}\;(\mathsf{Day}\;(\mathsf{Id}\;\Varid{x})\;\Varid{y}\;\Varid{f})\mathrel{=}\FN{fmap}\;(\Varid{f}\circ(\lambda \Varid{y}\to (\Varid{x},\Varid{y})))\;\Varid{y}{}\<[E]%
\\[\blanklineskip]%
\>[B]{}\rho^{-1}\mathbin{::}\mathsf{Functor}\;\Varid{f}\Rightarrow \Varid{f}\dayt\mathsf{Id}\xrightarrow{\sbt}\Varid{f}{}\<[E]%
\\
\>[B]{}\rho^{-1}\;(\mathsf{Day}\;\Varid{x}\;(\mathsf{Id}\;\Varid{y})\;\Varid{f})\mathrel{=}\FN{fmap}\;(\Varid{f}\circ(\lambda \Varid{z}\to (\Varid{z},\Varid{y})))\;\Varid{x}{}\<[E]%
\\[\blanklineskip]%
\>[B]{}\alpha^{-1}\mathbin{::}\Varid{f}\dayt(\Varid{g}\dayt\Varid{h})\xrightarrow{\sbt}(\Varid{f}\dayt\Varid{g})\dayt\Varid{h}{}\<[E]%
\\
\>[B]{}\alpha^{-1}\;(\mathsf{Day}\;\Varid{x}\;(\mathsf{Day}\;\Varid{y}\;\Varid{z}\;\Varid{f})\;\Varid{g})\mathrel{=}\mathsf{Day}\;(\mathsf{Day}\;\Varid{x}\;\Varid{y}\;\Varid{f}_{1})\;\Varid{z}\;\Varid{f}_{2}{}\<[E]%
\\
\>[B]{}\hsindent{21}{}\<[21]%
\>[21]{}\mathkw{where}\;{}\<[28]%
\>[28]{}\Varid{f}_{1}{}\<[32]%
\>[32]{}\mathrel{=}\lambda (\Varid{c},\Varid{e})\to (\Varid{c},\lambda \Varid{h}\to \Varid{f}\;(\Varid{e},\Varid{h})){}\<[E]%
\\
\>[28]{}\Varid{f}_{2}{}\<[32]%
\>[32]{}\mathrel{=}\lambda ((\Varid{c},\Varid{h}),\Varid{d})\to \Varid{g}\;(\Varid{c},\Varid{h}\;\Varid{d}){}\<[E]%
\ColumnHook
\end{hscode}\resethooks
}{We leave the definition of the inverses as an exercise.}
\end{coder}



\begin{mremark}[Alternative presentations of the Day convolution]
  In our setting of $\Set$ functors, the Day convolution has
  different alternative representations:
\begin{equation}
  \label{eq:alt_day}
  (F \dayt G)B \quad \cong \quad \int^A F A \times G(B^A) \quad \cong \quad \int^A F(B^A) \times G A
\end{equation}
\end{mremark}

\subsection{Monoids in $\EndD$}

A monoid in $\EndD$ amounts to:
\begin{itemize}
\item An endofunctor $F$, 
\item a natural transformation $m : F \dayt F \to F$,
\item and a unit $e : \Id \to F$; such that the following diagrams commute.

\[
\xymatrix@C+=10mm{
(F\dayt F)\dayt F\ar[rr]^{m \dayt F}& & F\dayt F \ar[d]^{m}\\
F\dayt (F\dayt F) \ar[u]^{\alpha} \ar[r]_-{F \dayt m}
& F\dayt F \ar[r]_-{m} & F \\
}\qquad
\xymatrix@C+=10mm{
    F \dayt F \ar[dr]^{m} 
 &
   F\dayt \Id \ar[l]_-{F \dayt e} \ar[d]^{\rho}
\\
   \Id\dayt F \ar[u]^-{e \dayt F} \ar[r]_{\lambda}
 &
   F
}
\]
\end{itemize}
From the unit $e$, one can consider the component $e_I : 1 \to F 1$. This
component defines a mapping which can be used as the unit morphism
for a lax monoidal functor. Similarly, using equation~\ref{eq:adj_day}, the morphism
$m : F \dayt F \to F$ is equivalent to a family of morphisms
\[ \vartheta(m)_{A,B} : F A \times F B \to F (A \times B) \]%
which is natural in $A$ and $B$. This family of morphisms corresponds to
the multiplicative transformation in a lax monoidal functor. Putting
together $F$, $\vartheta(m)$ and $e_I$, we obtain a strong lax monoidal functor on $\Set$,
that is, an applicative functor.




It remains to be seen if the converse is true: can a monoid in
$\EndD$ be defined from an applicative functor? Given
an applicative functor $(F, \phi, \eta)$, it easy to see that
a multiplication for the monoid can be given from $\phi$,
using equation~\ref{eq:adj_day} again.
What has to be seen it is if one can recover
the whole natural transformation $e : \Id \to F$ out of only one
component $\eta : 1 \to F 1$. We do so by using the strength
of $F$ (which exists since it is an endofunctor on $\Set$):
the following composition
\[
\xymatrix@C+=1.5cm{
A \ar[r]^{\langle \id, ! \rangle} & A \times 1 \ar[r]^{\id \times \eta} & A \times F 1 \ar[r]^{\funst_{A,1}} & F (A \times 1) \ar[r]^{F \pi_1} & F A
}
\]%
defines a morphism $e_A : A \to F A$ for each $A$.



All told, \emph{applicative functors are monoids in the category of
  endofunctors which is monoidal with respect to the Day convolution}.


\subsection{Exponential for Applicatives}
To apply the Cayley representation, first it must be determined if the
category $\EndD$ is monoidal closed.  To do so, we use the same
technique we used in section~\ref{sec:monad_exponential} for finding
the exponential of monads: we apply Yoneda
and then the universal property of exponentials.
%
\begin{align*}
G^F(B) &\cong \Nat{\Hom(B, -)}{G^F} \\
       &\cong \Nat{\Hom(B, -) \dayt F}{ G}
\end{align*}
Therefore, whenever the last expression makes sense, it can be used as
the definition of the exponential object. Since we are working on a
category of small functors, the expression always makes sense and the
exponential is always guaranteed to exist.
 Doing some further algebra,
an alternative form for $G^F$ can be derived~\cite{GoodDay}:
\[ G^F(B) \cong \Nat{F}{G (B \times -)} \]%
\begin{coder}
Using Haskell, this exponential can be represented as:

\begin{hscode}\SaveRestoreHook
\column{B}{@{}>{\hspre}l<{\hspost}@{}}%
\column{E}{@{}>{\hspre}l<{\hspost}@{}}%
\>[B]{}\mathkw{data}\;\mathsf{Exp}\;\Varid{f}\;\Varid{g}\;\Varid{b}\mathrel{=}\mathsf{Exp}\;(\forall \Varid{a}\hsforall .\ \Varid{f}\;\Varid{a}\to \Varid{g}\;(\Varid{b},\Varid{a})){}\<[E]%
\ColumnHook
\end{hscode}\resethooks

The components of the isomorphism showing it is an exponential are:
\begin{hscode}\SaveRestoreHook
\column{B}{@{}>{\hspre}l<{\hspost}@{}}%
\column{26}{@{}>{\hspre}l<{\hspost}@{}}%
\column{E}{@{}>{\hspre}l<{\hspost}@{}}%
\>[B]{}\varphi \mathbin{::}(\Varid{f}\dayt\Varid{g}\xrightarrow{\sbt}\Varid{h})\to \Varid{f}\xrightarrow{\sbt}\mathsf{Exp}\;\Varid{g}\;\Varid{h}{}\<[E]%
\\
\>[B]{}\varphi \;\Varid{m}\;\Varid{x}\mathrel{=}\mathsf{Exp}\;(\lambda \Varid{y}\to \Varid{m}\;(\mathsf{Day}\;\Varid{x}\;\Varid{y}\;\FN{id})){}\<[E]%
\\[\blanklineskip]%
\>[B]{}\varphi^{-1}\mathbin{::}\mathsf{Functor}\;\Varid{h}\Rightarrow (\Varid{f}\xrightarrow{\sbt}\mathsf{Exp}\;\Varid{g}\;\Varid{h})\to \Varid{f}\dayt\Varid{g}\xrightarrow{\sbt}\Varid{h}{}\<[E]%
\\
\>[B]{}\varphi^{-1}\;\Varid{f}\;(\mathsf{Day}\;\Varid{x}\;\Varid{y}\;\Varid{h})\mathrel{=}{}\<[26]%
\>[26]{}\FN{fmap}\;\Varid{h}\;(\Varid{t}\;\Varid{y}){}\<[E]%
\\
\>[26]{}\mathkw{where}\;\mathsf{Exp}\;\Varid{t}\mathrel{=}\Varid{f}\;\Varid{x}{}\<[E]%
\ColumnHook
\end{hscode}\resethooks
\end{coder}


We therefore conclude that $\EndD$ is a symmetric monoidal closed category.

\subsection{Free Applicatives}

By Proposition~\ref{prop:free}, the free monoid, viz. the
free applicative functor, exists.

\begin{coder}
  The direct application of proposition~\ref{prop:free} yields the
  following implementation of the free applicative functor.
\begin{hscode}\SaveRestoreHook
\column{B}{@{}>{\hspre}l<{\hspost}@{}}%
\column{E}{@{}>{\hspre}l<{\hspost}@{}}%
\>[B]{}\mathkw{data}\;\FN{Free}_{\dayt}\;\Varid{f}\;\Varid{a}\mathrel{=}\mathsf{Pure}\;\Varid{a}\mid \mathsf{Rec}\;((\Varid{f}\dayt\FN{Free}_{\dayt}\;\Varid{f})\;\Varid{a}){}\<[E]%
\ColumnHook
\end{hscode}\resethooks

Inlining the definition of \ensuremath{\dayt}, we obtain the simplified datatype

\begin{hscode}\SaveRestoreHook
\column{B}{@{}>{\hspre}l<{\hspost}@{}}%
\column{9}{@{}>{\hspre}l<{\hspost}@{}}%
\column{15}{@{}>{\hspre}l<{\hspost}@{}}%
\column{E}{@{}>{\hspre}l<{\hspost}@{}}%
\>[B]{}\mathkw{data}\;\FN{Free}_{\dayt}\;\Varid{f}\;\Varid{a}\;\mathkw{where}{}\<[E]%
\\
\>[B]{}\hsindent{9}{}\<[9]%
\>[9]{}\mathsf{Pure}{}\<[15]%
\>[15]{}\mathbin{::}\Varid{a}\to \FN{Free}_{\dayt}\;\Varid{f}\;\Varid{a}{}\<[E]%
\\
\>[B]{}\hsindent{9}{}\<[9]%
\>[9]{}\mathsf{Rec}{}\<[15]%
\>[15]{}\mathbin{::}\Varid{f}\;\Varid{c}\to \FN{Free}_{\dayt}\;\Varid{f}\;\Varid{d}\to ((\Varid{c},\Varid{d})\to \Varid{a})\to \FN{Free}_{\dayt}\;\Varid{f}\;\Varid{a}{}\<[E]%
\ColumnHook
\end{hscode}\resethooks

with the following instances:

\begin{hscode}\SaveRestoreHook
\column{B}{@{}>{\hspre}l<{\hspost}@{}}%
\column{3}{@{}>{\hspre}l<{\hspost}@{}}%
\column{16}{@{}>{\hspre}l<{\hspost}@{}}%
\column{23}{@{}>{\hspre}l<{\hspost}@{}}%
\column{24}{@{}>{\hspre}l<{\hspost}@{}}%
\column{E}{@{}>{\hspre}l<{\hspost}@{}}%
\>[B]{}\mathkw{instance}\;\mathsf{Functor}\;\Varid{f}\Rightarrow \mathsf{Functor}\;(\FN{Free}_{\dayt}\;\Varid{f})\;\mathkw{where}{}\<[E]%
\\
\>[B]{}\hsindent{3}{}\<[3]%
\>[3]{}\FN{fmap}\;\Varid{g}\;(\mathsf{Pure}\;\Varid{x}){}\<[23]%
\>[23]{}\mathrel{=}\mathsf{Pure}\;(\Varid{g}\;\Varid{x}){}\<[E]%
\\
\>[B]{}\hsindent{3}{}\<[3]%
\>[3]{}\FN{fmap}\;\Varid{g}\;(\mathsf{Rec}\;\Varid{x}\;\Varid{y}\;\Varid{f}){}\<[23]%
\>[23]{}\mathrel{=}\mathsf{Rec}\;\Varid{x}\;\Varid{y}\;(\Varid{g}\circ\Varid{f}){}\<[E]%
\\[\blanklineskip]%
\>[B]{}\mathkw{instance}\;\mathsf{Functor}\;\Varid{f}\Rightarrow \mathsf{Applicative}\;(\FN{Free}_{\dayt}\;\Varid{f})\;\mathkw{where}{}\<[E]%
\\
\>[B]{}\hsindent{3}{}\<[3]%
\>[3]{}\FN{pure}{}\<[24]%
\>[24]{}\mathrel{=}\mathsf{Pure}{}\<[E]%
\\
\>[B]{}\hsindent{3}{}\<[3]%
\>[3]{}\mathsf{Pure}\;\Varid{g}{}\<[16]%
\>[16]{}\circledast\Varid{z}{}\<[24]%
\>[24]{}\mathrel{=}\FN{fmap}\;\Varid{g}\;\Varid{z}{}\<[E]%
\\
\>[B]{}\hsindent{3}{}\<[3]%
\>[3]{}(\mathsf{Rec}\;\Varid{x}\;\Varid{y}\;\Varid{f}){}\<[16]%
\>[16]{}\circledast\Varid{z}{}\<[24]%
\>[24]{}\mathrel{=}\mathsf{Rec}\;\Varid{x}\;(\FN{pure}\;(,)\circledast\Varid{y}\circledast\Varid{z})\;(\lambda (\Varid{c},(\Varid{d},\Varid{a}))\to \Varid{f}\;(\Varid{c},\Varid{d})\;\Varid{a}){}\<[E]%
\ColumnHook
\end{hscode}\resethooks

There is no need to check that the instance satisfies the applicative laws
since the definition is derived from Proposition~\ref{prop:free}.

The implementation of the insertion of generators and the universal
morphism from the free applicative is:
\begin{hscode}\SaveRestoreHook
\column{B}{@{}>{\hspre}l<{\hspost}@{}}%
\column{21}{@{}>{\hspre}l<{\hspost}@{}}%
\column{E}{@{}>{\hspre}l<{\hspost}@{}}%
\>[B]{}\ins\mathbin{::}\mathsf{Functor}\;\Varid{a}\Rightarrow \Varid{a}\xrightarrow{\sbt}\FN{Free}_{\dayt}\;\Varid{a}{}\<[E]%
\\
\>[B]{}\ins\;\Varid{x}\mathrel{=}\mathsf{Rec}\;\Varid{x}\;(\mathsf{Pure}\;())\;\FN{fst}{}\<[E]%
\\[\blanklineskip]%
\>[B]{}\FN{free}\mathbin{::}(\mathsf{Functor}\;\Varid{a},\mathsf{Applicative}\;\Varid{b})\Rightarrow (\Varid{a}\xrightarrow{\sbt}\Varid{b})\to (\FN{Free}_{\dayt}\;\Varid{a}\xrightarrow{\sbt}\Varid{b}){}\<[E]%
\\
\>[B]{}\FN{free}\;\Varid{f}\;(\mathsf{Pure}\;\Varid{x}){}\<[21]%
\>[21]{}\mathrel{=}\FN{pure}\;\Varid{x}{}\<[E]%
\\
\>[B]{}\FN{free}\;\Varid{f}\;(\mathsf{Rec}\;\Varid{x}\;\Varid{y}\;\Varid{g}){}\<[21]%
\>[21]{}\mathrel{=}\FN{pure}\;(\FN{curry}\;\Varid{g})\circledast\Varid{f}\;\Varid{x}\circledast\FN{free}\;\Varid{f}\;\Varid{y}{}\<[E]%
\ColumnHook
\end{hscode}\resethooks

Alternative presentations of the Day convolution produce alternative
types for the free applicative. Using the two
alternative expressions for the Day convolution given in
equation~\ref{eq:alt_day}, we obtain two alternative
definitions of the free applicative functor:

\begin{hscode}\SaveRestoreHook
\column{B}{@{}>{\hspre}l<{\hspost}@{}}%
\column{7}{@{}>{\hspre}l<{\hspost}@{}}%
\column{17}{@{}>{\hspre}l<{\hspost}@{}}%
\column{48}{@{}>{\hspre}l<{\hspost}@{}}%
\column{E}{@{}>{\hspre}l<{\hspost}@{}}%
\>[B]{}\mathkw{data}\;\FN{Free}_{\dayt}^\prime\;\Varid{f}\;\Varid{a}\;\mathkw{where}{}\<[E]%
\\
\>[B]{}\hsindent{7}{}\<[7]%
\>[7]{}\mathsf{Pure'}{}\<[17]%
\>[17]{}\mathbin{::}\Varid{a}\to \FN{Free}_{\dayt}^\prime\;\Varid{f}\;\Varid{a}{}\<[E]%
\\
\>[B]{}\hsindent{7}{}\<[7]%
\>[7]{}\mathsf{Rec'}{}\<[17]%
\>[17]{}\mathbin{::}\Varid{f}\;\Varid{b}\to \FN{Free}_{\dayt}^\prime\;\Varid{f}\;(\Varid{b}\to \Varid{a}){}\<[48]%
\>[48]{}\to \FN{Free}_{\dayt}^\prime\;\Varid{f}\;\Varid{a}{}\<[E]%
\\[\blanklineskip]%
\>[B]{}\mathkw{data}\;\FN{Free}_{\dayt}^{\prime\prime}\;\Varid{f}\;\Varid{a}\;\mathkw{where}{}\<[E]%
\\
\>[B]{}\hsindent{7}{}\<[7]%
\>[7]{}\mathsf{Pure''}{}\<[17]%
\>[17]{}\mathbin{::}\Varid{a}\to \FN{Free}_{\dayt}^{\prime\prime}\;\Varid{f}\;\Varid{a}{}\<[E]%
\\
\>[B]{}\hsindent{7}{}\<[7]%
\>[7]{}\mathsf{Rec''}{}\<[17]%
\>[17]{}\mathbin{::}\Varid{f}\;(\Varid{b}\to \Varid{a})\to \FN{Free}_{\dayt}^{\prime\prime}\;\Varid{f}\;\Varid{b}\to \FN{Free}_{\dayt}^{\prime\prime}\;\Varid{f}\;\Varid{a}{}\<[E]%
\ColumnHook
\end{hscode}\resethooks
  

Hence, the two alternative presentations of the Day convolution given
in equation~\ref{eq:alt_day} give rise to the two notions of free
applicative functor found by Capriotti and
Kaposi~\shortcite{Capriotti2014}.
\end{coder}

\subsection{Cayley Representation for Applicatives}

Having found the exponential for applicatives, we may apply
theorem~\ref{thm:Cayley} and construct the corresponding Cayley
representation.

\begin{coder}
The Cayley representation is the exponential of a functor over itself.

\begin{hscode}\SaveRestoreHook
\column{B}{@{}>{\hspre}l<{\hspost}@{}}%
\column{3}{@{}>{\hspre}l<{\hspost}@{}}%
\column{19}{@{}>{\hspre}l<{\hspost}@{}}%
\column{20}{@{}>{\hspre}l<{\hspost}@{}}%
\column{23}{@{}>{\hspre}l<{\hspost}@{}}%
\column{38}{@{}>{\hspre}l<{\hspost}@{}}%
\column{E}{@{}>{\hspre}l<{\hspost}@{}}%
\>[B]{}\mathkw{type}\;\mathsf{Rep}\;\Varid{f}\mathrel{=}\mathsf{Exp}\;\Varid{f}\;\Varid{f}{}\<[E]%
\\[\blanklineskip]%
\>[B]{}\mathkw{instance}\;\mathsf{Functor}\;\Varid{f}\Rightarrow \mathsf{Functor}\;(\mathsf{Rep}\;\Varid{f})\;\mathkw{where}{}\<[E]%
\\
\>[B]{}\hsindent{3}{}\<[3]%
\>[3]{}\FN{fmap}\;\Varid{f}\;(\mathsf{Exp}\;\Varid{h})\mathrel{=}\mathsf{Exp}\;(\FN{fmap}\;(\lambda (\Varid{x},\Varid{y})\to (\Varid{f}\;\Varid{x},\Varid{y}))\circ\Varid{h}){}\<[E]%
\\[\blanklineskip]%
\>[B]{}\mathkw{instance}\;\mathsf{Functor}\;\Varid{f}\Rightarrow \mathsf{Applicative}\;(\mathsf{Rep}\;\Varid{f})\;\mathkw{where}{}\<[E]%
\\
\>[B]{}\hsindent{3}{}\<[3]%
\>[3]{}\FN{pure}\;\Varid{c}{}\<[20]%
\>[20]{}\mathrel{=}{}\<[23]%
\>[23]{}\mathsf{Exp}\;(\FN{fmap}\;(\Varid{c},)){}\<[E]%
\\
\>[B]{}\hsindent{3}{}\<[3]%
\>[3]{}\mathsf{Exp}\;\Varid{f}\circledast\mathsf{Exp}\;\Varid{a}{}\<[20]%
\>[20]{}\mathrel{=}\mathsf{Exp}\;(\FN{fmap}\;\Varid{g}\circ\Varid{a}\circ\Varid{f}){}\<[E]%
\\
\>[3]{}\hsindent{16}{}\<[19]%
\>[19]{}\mathkw{where}\;\Varid{g}\;(\Varid{x},(\Varid{f},\Varid{c})){}\<[38]%
\>[38]{}\mathrel{=}(\Varid{f}\;\Varid{x},\Varid{c}){}\<[E]%
\ColumnHook
\end{hscode}\resethooks

Again, there is no need to check compliance with applicative laws
because the instance is derived from the general construction of the
monoid of endomorphism.

Finally, from theorem~\ref{thm:Cayley}, we obtain the applicative
morphism \ensuremath{\rep} and the natural transformation \ensuremath{\abs}, together with
the property that \ensuremath{\abs\circ\rep\mathrel{=}\FN{id}}.
\begin{hscode}\SaveRestoreHook
\column{B}{@{}>{\hspre}l<{\hspost}@{}}%
\column{15}{@{}>{\hspre}l<{\hspost}@{}}%
\column{E}{@{}>{\hspre}l<{\hspost}@{}}%
\>[B]{}\rep\mathbin{::}\mathsf{Applicative}\;\Varid{f}\Rightarrow \Varid{f}\xrightarrow{\sbt}\mathsf{Rep}\;\Varid{f}{}\<[E]%
\\
\>[B]{}\rep\;\Varid{x}\mathrel{=}\mathsf{Exp}\;(\lambda \Varid{y}\to \FN{pure}\;(,)\circledast\Varid{x}\circledast\Varid{y}){}\<[E]%
\\[\blanklineskip]%
\>[B]{}\abs\mathbin{::}\mathsf{Applicative}\;\Varid{f}\Rightarrow \mathsf{Rep}\;\Varid{f}\xrightarrow{\sbt}\Varid{f}{}\<[E]%
\\
\>[B]{}\abs\;(\mathsf{Exp}\;\Varid{t})\mathrel{=}{}\<[15]%
\>[15]{}\FN{fmap}\;\FN{fst}\;(\Varid{t}\;(\FN{pure}\;())){}\<[E]%
\ColumnHook
\end{hscode}\resethooks
\vspace{-5mm}
\end{coder}


%

\section{Pre-Arrows as Monoids}
\label{sec:prearrows}

Having successfully applied the Cayley representation to monads and
applicatives, we wonder if we can find a representation for a third
popular notion of computation: arrows. Arrows~\cite{Hughes-SCP00} were
already studied as monoids~\cite{jacobs2009categorical}, resulting in
a monoid in the category of profunctors. We briefly review these
results.

A profunctor from $\C$ to $\D$ is a functor $\D^{\op} \times \C
\to \Set$, sometimes written as $\C \profunctor \D$.  In a sense,
functors are to functions what profunctors are to relations.  A
morphism between two profunctors is a natural transformation between
their underlying functors.

\begin{coder}
  We indicate that a type constructor \ensuremath{\Varid{h}\mathbin{::}\mathbin{*}\to \mathbin{*}\to \mathbin{*}} is a
  profunctor by providing an instance of the following type class.

\begin{hscode}\SaveRestoreHook
\column{B}{@{}>{\hspre}l<{\hspost}@{}}%
\column{3}{@{}>{\hspre}l<{\hspost}@{}}%
\column{E}{@{}>{\hspre}l<{\hspost}@{}}%
\>[B]{}\mathkw{class}\;\mathsf{Profunctor}\;\Varid{h}\;\mathkw{where}{}\<[E]%
\\
\>[B]{}\hsindent{3}{}\<[3]%
\>[3]{}\FN{dimap}\mathbin{::}(\Varid{d'}\to \Varid{d})\to (\Varid{c}\to \Varid{c'})\to \Varid{h}\;\Varid{d}\;\Varid{c}\to \Varid{h}\;\Varid{d'}\;\Varid{c'}{}\<[E]%
\ColumnHook
\end{hscode}\resethooks

such that the following laws hold
\begin{align*}
\ensuremath{\FN{dimap}\;\FN{id}\;\FN{id}} &= \ensuremath{\FN{id}} \\
\ensuremath{\FN{dimap}\;(\Varid{f}\circ\Varid{g})\;(\Varid{h}\circ\Varid{i})} &= \ensuremath{\FN{dimap}\;\Varid{g}\;\Varid{h}\circ\FN{dimap}\;\Varid{f}\;\Varid{i}}
\end{align*}
Notice how, as opposed to a bifunctor, the type constructor is
contravariant on its first argument.
\end{coder}

\begin{mdefinition}
  The category of profunctors from $\C$ to $\D$, denoted
  $\Prof{\C}{\D}$, has as objects profunctors from $\C$ to $\D$, and
  as morphisms natural transformation between functors $\D^{\op}
  \times \C \to\Set$.
\end{mdefinition}

From now on, we will focus on profunctors $\C\profunctor\C$, where
$\C$ is a small cartesian closed subcategory of $\Set$ with inclusion
$J : \C \to \Set$.  To avoid notational clutter, we omit the functor
$J$ when considering elements of $\C$ as elements of $\Set$.

%
Profunctors can be composed in such a
way that give a notion of tensor~\cite{Benabou1973}. Given
two profunctors $F, G : \C \profunctor \C$, their composition is
\[ (F \oX G)(A,B) = \int^{Z} F(A,Z) \times G(Z, B) \]%
\begin{coder}
The tensor is implemented in Haskell as follows:
\begin{hscode}\SaveRestoreHook
\column{B}{@{}>{\hspre}l<{\hspost}@{}}%
\column{3}{@{}>{\hspre}l<{\hspost}@{}}%
\column{E}{@{}>{\hspre}l<{\hspost}@{}}%
\>[B]{}\mathkw{data}\;(\otimes)\;\Varid{f}\;\Varid{g}\;\Varid{a}\;\Varid{b}\mathrel{=}\forall \Varid{z}\hsforall .\ (\Varid{f}\;\Varid{a}\;\Varid{z})\otimes(\Varid{g}\;\Varid{z}\;\Varid{b}){}\<[E]%
\\[\blanklineskip]%
\>[B]{}\mathkw{instance}\;(\mathsf{Profunctor}\;\Varid{f},\mathsf{Profunctor}\;\Varid{g})\Rightarrow \mathsf{Profunctor}\;(\Varid{f}\otimes\Varid{g})\;\mathkw{where}{}\<[E]%
\\
\>[B]{}\hsindent{3}{}\<[3]%
\>[3]{}\FN{dimap}\;\Varid{m}_{1}\;\Varid{m}_{2}\;(\Varid{f}\otimes\Varid{g})\mathrel{=}(\FN{dimap}\;\Varid{m}_{1}\;\FN{id}\;\Varid{f})\otimes(\FN{dimap}\;\FN{id}\;\Varid{m}_{2}\;\Varid{g}){}\<[E]%
\ColumnHook
\end{hscode}\resethooks

  
\end{coder}

The functor $\Hom : \C^{\op} \times \C \to \Set$ is small
and it is the unit for the composition:
\[ (F \oX \Hom)(A,B) = \int^{P} F(A,P) \times \Hom(P,B) \cong F(A,B) \]%
where the isomorphism holds by the Yoneda lemma. This calculation
is used to define a natural isomorphism $\rho : F \oX \Hom \cong F$.
Likewise, natural isomorphisms $\lambda : \Hom \oX F \cong F$ and
$\alpha : (F \oX G) \oX H \cong F \oX (G \oX H)$
can be defined.

\begin{coder}
We represent morphisms between profunctors as
\begin{hscode}\SaveRestoreHook
\column{B}{@{}>{\hspre}l<{\hspost}@{}}%
\column{E}{@{}>{\hspre}l<{\hspost}@{}}%
\>[B]{}\mathkw{type}\;\Varid{f}\xrightarrow{\sbt\ \sbt}\Varid{g}\mathrel{=}\forall \Varid{a}\hsforall \;\Varid{b}.\ \Varid{f}\;\Varid{a}\;\Varid{b}\to \Varid{g}\;\Varid{a}\;\Varid{b}{}\<[E]%
\ColumnHook
\end{hscode}\resethooks

The implementation of $\lambda$, $\rho$, and $\alpha$ are as follows:

 \begin{hscode}\SaveRestoreHook
\column{B}{@{}>{\hspre}l<{\hspost}@{}}%
\column{7}{@{}>{\hspre}l<{\hspost}@{}}%
\column{15}{@{}>{\hspre}l<{\hspost}@{}}%
\column{23}{@{}>{\hspre}l<{\hspost}@{}}%
\column{24}{@{}>{\hspre}l<{\hspost}@{}}%
\column{E}{@{}>{\hspre}l<{\hspost}@{}}%
\>[B]{}\mathkw{type}\;\mathsf{Hom}\mathrel{=}(\to ){}\<[E]%
\\[\blanklineskip]%
\>[B]{}\lambda \mathbin{::}\mathsf{Profunctor}\;\Varid{f}\Rightarrow \mathsf{Hom}\otimes\Varid{f}\xrightarrow{\sbt\ \sbt}\Varid{f}{}\<[E]%
\\
\>[B]{}\lambda \;(\Varid{f}\otimes\Varid{x}){}\<[23]%
\>[23]{}\mathrel{=}\FN{dimap}\;\Varid{f}\;\FN{id}\;\Varid{x}{}\<[E]%
\\[\blanklineskip]%
\>[B]{}\rho {}\<[7]%
\>[7]{}\mathbin{::}\mathsf{Profunctor}\;\Varid{f}\Rightarrow \Varid{f}\otimes\mathsf{Hom}\xrightarrow{\sbt\ \sbt}\Varid{f}{}\<[E]%
\\
\>[B]{}\rho \;(\Varid{x}\otimes\Varid{f}){}\<[15]%
\>[15]{}\mathrel{=}\FN{dimap}\;\FN{id}\;\Varid{f}\;\Varid{x}{}\<[E]%
\\[\blanklineskip]%
\>[B]{}\alpha \mathbin{::}(\Varid{f}\otimes\Varid{g})\otimes\Varid{h}\xrightarrow{\sbt\ \sbt}\Varid{f}\otimes(\Varid{g}\otimes\Varid{h}){}\<[E]%
\\
\>[B]{}\alpha \;((\Varid{f}\otimes\Varid{g})\otimes\Varid{h}){}\<[24]%
\>[24]{}\mathrel{=}\Varid{f}\otimes(\Varid{g}\otimes\Varid{h}){}\<[E]%
\ColumnHook
\end{hscode}\resethooks
\end{coder}
\noindent Thus, a monoidal structure can be given for 
$\left[ \C^\op\times \C,\Set\right]$, with
composition $\oX$ as its tensor, and $\Hom$ as its unit. 
We denote this monoidal category by $\Pro$.

Which are the monoids in this monoidal category?
A monoid in $\Pro$ amounts to:
\begin{itemize}
\item A profunctor $A$, 
\item a natural transformation $m : A \oX A \to A$,
\item and a unit $e : \Hom \to A$; such that the diagrams
\[
\xymatrix@C+=10mm{
(A\oX A)\oX A\ar[rr]^{m \oX A}& & A\oX A \ar[d]^{m}\\
A\oX (A\oX A) \ar[u]^{\alpha} \ar[r]_-{A \oX m}
& A\oX F \ar[r]_-{m} & A \\
}\qquad
\xymatrix@C+=10mm{
    A \oX A \ar[dr]^{m} 
 &
   A\oX \Hom \ar[l]_-{A \oX e} \ar[d]^{\rho}
\\
   \Hom\oX A \ar[u]^-{e \oX A} \ar[r]_{\lambda}
 &
   A
}
\]
commute.
\end{itemize}


Using the isomorphism
\[ \left( \int^Z A(X,Z) \times A(Z,Y) \right) \to A(X,Y) \cong
\int_Z A(X,Z) \times A(Z,Y) \to A(X,Y) \]%
we get that a natural transformation $m : A \oX A \to A$ is equivalent to a family
of morphisms $m_{X,Y,Z} : A(X,Z) \times A(Z, Y) \to A(X,Y)$ which is natural in
$X$ and $Y$ and dinatural in $Z$.

This presentation makes the connection with arrows evident: $m$
corresponds to the operator \ensuremath{(\ensuremath{\ggg})} and $e$ corresponds to
\ensuremath{\FN{arr}}. Unfortunately, the \ensuremath{\FN{first}} operation is missing.  We postpone this
problem until the next section, and in the remainder of this section
focus on monoids in $\Pro$, i.e. arrows without a
\ensuremath{\FN{first}} operation, which we call \emph{pre-arrows}.

\begin{coder}
We introduce a class to represent the monoids in this category. It is simply
a restriction of the \ensuremath{\mathsf{Arrow}} class, omitting the \ensuremath{\FN{first}} operation.
\begin{hscode}\SaveRestoreHook
\column{B}{@{}>{\hspre}l<{\hspost}@{}}%
\column{3}{@{}>{\hspre}l<{\hspost}@{}}%
\column{10}{@{}>{\hspre}l<{\hspost}@{}}%
\column{E}{@{}>{\hspre}l<{\hspost}@{}}%
\>[B]{}\mathkw{class}\;\mathsf{Profunctor}\;\Varid{a}\Rightarrow \mathsf{PreArrow}\;\Varid{a}\;\mathkw{where}{}\<[E]%
\\
\>[B]{}\hsindent{3}{}\<[3]%
\>[3]{}\FN{arr}{}\<[10]%
\>[10]{}\mathbin{::}(\Varid{b}\to \Varid{c})\to \Varid{a}\;\Varid{b}\;\Varid{c}{}\<[E]%
\\
\>[B]{}\hsindent{3}{}\<[3]%
\>[3]{}(\ensuremath{\ggg}){}\<[10]%
\>[10]{}\mathbin{::}\Varid{a}\;\Varid{b}\;\Varid{c}\to \Varid{a}\;\Varid{c}\;\Varid{d}\to \Varid{a}\;\Varid{b}\;\Varid{d}{}\<[E]%
\ColumnHook
\end{hscode}\resethooks

The laws that must hold are
\begin{align*}
\ensuremath{(\Varid{a}\ensuremath{\ggg}\Varid{b})\ensuremath{\ggg}\Varid{c}} &= \ensuremath{\Varid{a}\ensuremath{\ggg}(\Varid{b}\ensuremath{\ggg}\Varid{c})} \\
\ensuremath{\FN{arr}\;\Varid{f}\ensuremath{\ggg}\Varid{a}} &= \ensuremath{\FN{dimap}\;\Varid{f}\;\FN{id}\;\Varid{a}} \\
\ensuremath{\Varid{a}\ensuremath{\ggg}\FN{arr}\;\Varid{f}} &= \ensuremath{\FN{dimap}\;\FN{id}\;\Varid{f}\;\Varid{a}} \\
\ensuremath{\FN{arr}\;(\Varid{g}\circ\Varid{f})} &= \ensuremath{\FN{arr}\;\Varid{f}\ensuremath{\ggg}\FN{arr}\;\Varid{g}}
\end{align*}
\end{coder}


\subsection{Exponential for Pre-Arrows}
\label{sec:exponential_prearrows}

\extended[The exponential in $\Pro$ exists~\cite{Benabou1973} and a simple calculation using the Yoneda Lemma shows it to be 
$$B^A(X,Y) = \Nat{A(Y,-)}{B(X,-)}.$$ 
]
{
Exponential objects in $\Pro$ exist~\cite{Benabou1973} and can be calculated
with the help of the Yoneda lemma.%
 The calculation of the exponential
of $B$ to $A$ is:
\begin{align*}
B^A(X,Y) &\cong \Nat{\Hom((X,Y), -)}{B^A} \\
         &\cong \Nat{\Hom((X,Y), -) \oX A}{B} \\
         &\cong \int_{C,D} (\Hom((X,Y), -) \oX A)(C,D) \to B(C,D) \\
         &\cong \int_{C,D} \left[ \int^P \Hom((X,Y),(C,P)) \times A(P,D) \right] \to B(C,D) \\
         &\cong \int_{C,D}  \left[ \int^P \Hom(X,C) \times \Hom(Y,P) \times A(P,D) \right] \to B(C,D) \\
         &\cong \int_{C,D} \Hom(X,C) \times \left[ \int^P \Hom(Y,P) \times A(P,D) \right] \to B(C,D) \\
         &\cong \int_{C,D} \Hom(X,C) \times A(Y,D) \to B(C,D) \\
         &\cong \int_{C,D} A(Y,D) \times \Hom(X,C) \to B(C,D) \\
         &\cong \int_{D}  A(Y,D) \to \int_{C} \Hom(X,C) \to B(C,D) \\
         &\cong \int_{D}  A(Y,D) \to B(X,D) \\
         &\cong \Nat{A(Y,-)}{B(X,-)}
\end{align*}
}%
\begin{coder}
The implementation of exponentials in $\Pro$ follows the definition above:
\begin{hscode}\SaveRestoreHook
\column{B}{@{}>{\hspre}l<{\hspost}@{}}%
\column{3}{@{}>{\hspre}l<{\hspost}@{}}%
\column{E}{@{}>{\hspre}l<{\hspost}@{}}%
\>[B]{}\mathkw{data}\;\mathsf{Exp}\;\Varid{a}\;\Varid{b}\;\Varid{x}\;\Varid{y}\mathrel{=}\mathsf{Exp}\;(\forall \Varid{d}\hsforall .\ \Varid{a}\;\Varid{y}\;\Varid{d}\to \Varid{b}\;\Varid{x}\;\Varid{d}){}\<[E]%
\\[\blanklineskip]%
\>[B]{}\mathkw{instance}\;(\mathsf{Profunctor}\;\Varid{g},\mathsf{Profunctor}\;\Varid{h})\Rightarrow \mathsf{Profunctor}\;(\mathsf{Exp}\;\Varid{g}\;\Varid{h})\;\mathkw{where}{}\<[E]%
\\
\>[B]{}\hsindent{3}{}\<[3]%
\>[3]{}\FN{dimap}\;\Varid{m}_{1}\;\Varid{m}_{2}\;(\mathsf{Exp}\;\Varid{gh})\mathrel{=}\mathsf{Exp}\;(\FN{dimap}\;\Varid{m}_{1}\;\FN{id}\circ\Varid{gh}\circ\FN{dimap}\;\Varid{m}_{2}\;\FN{id}){}\<[E]%
\ColumnHook
\end{hscode}\resethooks

The components of the isomorphism which shows that \ensuremath{\mathsf{Exp}} is an
exponential are:

\begin{hscode}\SaveRestoreHook
\column{B}{@{}>{\hspre}l<{\hspost}@{}}%
\column{7}{@{}>{\hspre}l<{\hspost}@{}}%
\column{16}{@{}>{\hspre}l<{\hspost}@{}}%
\column{19}{@{}>{\hspre}l<{\hspost}@{}}%
\column{E}{@{}>{\hspre}l<{\hspost}@{}}%
\>[B]{}\varphi {}\<[16]%
\>[16]{}\mathbin{::}(\Varid{f}\otimes\Varid{g}\xrightarrow{\sbt\ \sbt}\Varid{h})\to (\Varid{f}\xrightarrow{\sbt\ \sbt}\mathsf{Exp}\;\Varid{g}\;\Varid{h}){}\<[E]%
\\
\>[B]{}\varphi \;{}\<[7]%
\>[7]{}\Varid{m}\;\Varid{f}{}\<[19]%
\>[19]{}\mathrel{=}\mathsf{Exp}\;(\lambda \Varid{g}\to \Varid{m}\;(\Varid{f}\otimes\Varid{g})){}\<[E]%
\\[\blanklineskip]%
\>[B]{}\varphi^{-1}{}\<[16]%
\>[16]{}\mathbin{::}(\Varid{f}\xrightarrow{\sbt\ \sbt}\mathsf{Exp}\;\Varid{g}\;\Varid{h})\to (\Varid{f}\otimes\Varid{g}\xrightarrow{\sbt\ \sbt}\Varid{h}){}\<[E]%
\\
\>[B]{}\varphi^{-1}\;{}\<[7]%
\>[7]{}\Varid{m}\;(\Varid{f}\otimes\Varid{g}){}\<[19]%
\>[19]{}\mathrel{=}\Varid{e}\;\Varid{g}\;\mathkw{where}\;\mathsf{Exp}\;\Varid{e}\mathrel{=}\Varid{m}\;\Varid{f}{}\<[E]%
\ColumnHook
\end{hscode}\resethooks
\vspace{-5mm}
\end{coder}

\subsection{Free Pre-Arrows}
By Proposition~\ref{prop:free}, the free monoid, viz. the
free pre-arrow, exists.
\begin{coder}
  The direct application of Proposition~\ref{prop:free} yields the
  following implementation of the free pre-arrow.

\begin{hscode}\SaveRestoreHook
\column{B}{@{}>{\hspre}l<{\hspost}@{}}%
\column{9}{@{}>{\hspre}l<{\hspost}@{}}%
\column{16}{@{}>{\hspre}l<{\hspost}@{}}%
\column{20}{@{}>{\hspre}l<{\hspost}@{}}%
\column{E}{@{}>{\hspre}l<{\hspost}@{}}%
\>[B]{}\mathkw{data}\;\FN{Free}_{\oX}\;\Varid{a}\;\Varid{x}\;\Varid{y}\;\mathkw{where}{}\<[E]%
\\
\>[B]{}\hsindent{9}{}\<[9]%
\>[9]{}\mathsf{Hom}{}\<[16]%
\>[16]{}\mathbin{::}(\Varid{x}\to \Varid{y})\to \FN{Free}_{\oX}\;\Varid{a}\;\Varid{x}\;\Varid{y}{}\<[E]%
\\
\>[B]{}\hsindent{9}{}\<[9]%
\>[9]{}\mathsf{Comp}{}\<[16]%
\>[16]{}\mathbin{::}{}\<[20]%
\>[20]{}\Varid{a}\;\Varid{x}\;\Varid{p}\to \FN{Free}_{\oX}\;\Varid{a}\;\Varid{p}\;\Varid{y}\to \FN{Free}_{\oX}\;\Varid{a}\;\Varid{x}\;\Varid{y}{}\<[E]%
\ColumnHook
\end{hscode}\resethooks

with the following instances:

\begin{hscode}\SaveRestoreHook
\column{B}{@{}>{\hspre}l<{\hspost}@{}}%
\column{3}{@{}>{\hspre}l<{\hspost}@{}}%
\column{15}{@{}>{\hspre}l<{\hspost}@{}}%
\column{25}{@{}>{\hspre}l<{\hspost}@{}}%
\column{E}{@{}>{\hspre}l<{\hspost}@{}}%
\>[B]{}\mathkw{instance}\;\mathsf{Profunctor}\;\Varid{a}\Rightarrow \mathsf{Profunctor}\;(\FN{Free}_{\oX}\;\Varid{a})\;\mathkw{where}{}\<[E]%
\\
\>[B]{}\hsindent{3}{}\<[3]%
\>[3]{}\FN{dimap}\;\Varid{f}\;\Varid{g}\;(\mathsf{Hom}\;\Varid{h}){}\<[25]%
\>[25]{}\mathrel{=}\mathsf{Hom}\;(\Varid{g}\circ\Varid{h}\circ\Varid{f}){}\<[E]%
\\
\>[B]{}\hsindent{3}{}\<[3]%
\>[3]{}\FN{dimap}\;\Varid{f}\;\Varid{g}\;(\mathsf{Comp}\;\Varid{x}\;\Varid{y}){}\<[25]%
\>[25]{}\mathrel{=}\mathsf{Comp}\;(\FN{dimap}\;\Varid{f}\;\FN{id}\;\Varid{x})\;(\FN{dimap}\;\FN{id}\;\Varid{g}\;\Varid{y}){}\<[E]%
\\[\blanklineskip]%
\>[B]{}\mathkw{instance}\;\mathsf{Profunctor}\;\Varid{a}\Rightarrow \mathsf{PreArrow}\;(\FN{Free}_{\oX}\;\Varid{a})\;\mathkw{where}{}\<[E]%
\\
\>[B]{}\hsindent{3}{}\<[3]%
\>[3]{}\FN{arr}\;\Varid{f}{}\<[25]%
\>[25]{}\mathrel{=}\mathsf{Hom}\;\Varid{f}{}\<[E]%
\\
\>[B]{}\hsindent{3}{}\<[3]%
\>[3]{}(\mathsf{Hom}\;\Varid{f}){}\<[15]%
\>[15]{}\ensuremath{\ggg}\Varid{c}{}\<[25]%
\>[25]{}\mathrel{=}\FN{dimap}\;\Varid{f}\;\FN{id}\;\Varid{c}{}\<[E]%
\\
\>[B]{}\hsindent{3}{}\<[3]%
\>[3]{}(\mathsf{Comp}\;\Varid{x}\;\Varid{y}){}\<[15]%
\>[15]{}\ensuremath{\ggg}\Varid{c}{}\<[25]%
\>[25]{}\mathrel{=}\mathsf{Comp}\;\Varid{x}\;(\Varid{y}\ensuremath{\ggg}\Varid{c}){}\<[E]%
\ColumnHook
\end{hscode}\resethooks
There is no need to check that the instance satisfies the pre-arrow laws
since the definition is derived from Proposition~\ref{prop:free}.

The insertion of generators and the universal morphism from the free
pre-arrow are:
\begin{hscode}\SaveRestoreHook
\column{B}{@{}>{\hspre}l<{\hspost}@{}}%
\column{20}{@{}>{\hspre}l<{\hspost}@{}}%
\column{E}{@{}>{\hspre}l<{\hspost}@{}}%
\>[B]{}\ins\mathbin{::}\mathsf{Profunctor}\;\Varid{a}\Rightarrow \Varid{a}\xrightarrow{\sbt\ \sbt}\FN{Free}_{\oX}\;\Varid{a}{}\<[E]%
\\
\>[B]{}\ins\;\Varid{x}\mathrel{=}\mathsf{Comp}\;\Varid{x}\;(\FN{arr}\;\FN{id}){}\<[E]%
\\[\blanklineskip]%
\>[B]{}\FN{free}\mathbin{::}(\mathsf{Profunctor}\;\Varid{a},\mathsf{PreArrow}\;\Varid{b})\Rightarrow (\Varid{a}\xrightarrow{\sbt\ \sbt}\Varid{b})\to (\FN{Free}_{\oX}\;\Varid{a}\xrightarrow{\sbt\ \sbt}\Varid{b}){}\<[E]%
\\
\>[B]{}\FN{free}\;\Varid{f}\;(\mathsf{Hom}\;\Varid{g}){}\<[20]%
\>[20]{}\mathrel{=}\FN{arr}\;\Varid{g}{}\<[E]%
\\
\>[B]{}\FN{free}\;\Varid{f}\;(\mathsf{Comp}\;\Varid{x}\;\Varid{y}){}\<[20]%
\>[20]{}\mathrel{=}\Varid{f}\;\Varid{x}\ensuremath{\ggg}\FN{free}\;\Varid{f}\;\Varid{y}{}\<[E]%
\ColumnHook
\end{hscode}\resethooks
\end{coder}

\subsection{Cayley Representation of Pre-Arrows}

Having found the exponential for pre-arrows, we may apply
theorem~\ref{thm:Cayley} and construct the corresponding Cayley
representation.

\begin{coder}
The Cayley representation is the exponential of a profunctor over itself.

\begin{hscode}\SaveRestoreHook
\column{B}{@{}>{\hspre}l<{\hspost}@{}}%
\column{3}{@{}>{\hspre}l<{\hspost}@{}}%
\column{12}{@{}>{\hspre}l<{\hspost}@{}}%
\column{25}{@{}>{\hspre}l<{\hspost}@{}}%
\column{E}{@{}>{\hspre}l<{\hspost}@{}}%
\>[B]{}\mathkw{type}\;\mathsf{Rep}\;\Varid{a}\mathrel{=}\mathsf{Exp}\;\Varid{a}\;\Varid{a}{}\<[E]%
\\[\blanklineskip]%
\>[B]{}\mathkw{instance}\;\mathsf{Profunctor}\;\Varid{a}\Rightarrow \mathsf{PreArrow}\;(\mathsf{Rep}\;\Varid{a})\;\mathkw{where}{}\<[E]%
\\
\>[B]{}\hsindent{3}{}\<[3]%
\>[3]{}\FN{arr}\;\Varid{f}{}\<[25]%
\>[25]{}\mathrel{=}\mathsf{Exp}\;(\lambda \Varid{y}\to \FN{dimap}\;\Varid{f}\;\FN{id}\;\Varid{y}){}\<[E]%
\\
\>[B]{}\hsindent{3}{}\<[3]%
\>[3]{}(\mathsf{Exp}\;\Varid{f}){}\<[12]%
\>[12]{}\ensuremath{\ggg}(\mathsf{Exp}\;\Varid{g}){}\<[25]%
\>[25]{}\mathrel{=}\mathsf{Exp}\;(\lambda \Varid{y}\to \Varid{f}\;(\Varid{g}\;\Varid{y})){}\<[E]%
\ColumnHook
\end{hscode}\resethooks

Again, there is no need to check compliance with pre-arrow laws
because the instance is derived from the general construction of the
monoid of endomorphism.

Finally, from theorem~\ref{thm:Cayley}, we obtain the pre-arrow
morphism \ensuremath{\rep} and the natural transformation \ensuremath{\abs}, together with
the property that \ensuremath{\abs\circ\rep\mathrel{=}\FN{id}}.

\begin{hscode}\SaveRestoreHook
\column{B}{@{}>{\hspre}l<{\hspost}@{}}%
\column{23}{@{}>{\hspre}l<{\hspost}@{}}%
\column{25}{@{}>{\hspre}l<{\hspost}@{}}%
\column{E}{@{}>{\hspre}l<{\hspost}@{}}%
\>[B]{}\rep\mathbin{::}\mathsf{PreArrow}\;\Varid{a}\Rightarrow \Varid{a}\xrightarrow{\sbt\ \sbt}\mathsf{Rep}\;\Varid{a}{}\<[E]%
\\
\>[B]{}\rep\;\Varid{x}{}\<[23]%
\>[23]{}\mathrel{=}\mathsf{Exp}\;(\lambda \Varid{y}\to \Varid{x}\ensuremath{\ggg}\Varid{y}){}\<[E]%
\\[\blanklineskip]%
\>[B]{}\abs\mathbin{::}\mathsf{PreArrow}\;\Varid{a}\Rightarrow \mathsf{Rep}\;\Varid{a}\xrightarrow{\sbt\ \sbt}\Varid{a}{}\<[E]%
\\
\>[B]{}\abs\;(\mathsf{Exp}\;\Varid{f}){}\<[25]%
\>[25]{}\mathrel{=}\Varid{f}\;(\FN{arr}\;\FN{id}){}\<[E]%
\ColumnHook
\end{hscode}\resethooks
\vspace{-5mm}
\end{coder}

%

\section{Arrows as Monoids}
\label{sec:arrows}
Returning to the problem of arrows as monoids, we need to internalise
the \ensuremath{\FN{first}} operation in the categorical presentation. Jacobs et
al.~\shortcite{jacobs2009categorical} solve this problem by adjoining
an \ensuremath{\Varid{ist}} operator to monoids in $\Pro$: an arrow is a monoid $(A, m,
e)$ together with a family of morphisms $ist : A(X,Y) \to A(X,Y\times
X)$.  We take an alternative path. We work on a category of strong
profunctors (profunctors with a \ensuremath{\FN{first}}-like operator), and then
consider monoids in this new monoidal category.


\begin{mdefinition}
  A \emph{strength} for a profunctor $F : \C^{\op} \times \C \to \Set$
  is a family of morphisms
\[ \sst_{X,Y,Z} : F(X,Y) \to F(X \times Z, Y \times Z) \]%
that is natural in $X$, $Y$ and dinatural in $Z$, such that the following
diagrams commute.

\[
\xymatrix@C+=2cm{
F(X,Y) \ar[d]_{\sst_1} \ar[rd]^{F(\pi_1, \id)} \\
F(X\times 1, Y\times 1) \ar[r]_{F(\id, \pi_1)} & F(X\times 1, Y) \\
}
\]
\[
\xymatrix@C+=2cm{
F(X,Y) \ar[d]_{\sst_{V\times W}} \ar[r]^{\sst_V} & F(X\times V, Y \times V) \ar[d]^{\sst_W} \\
F(X\times (V \times W), Y\times (V \times W)) \ar[r]_{F(\alpha, \alpha^{-1})} & F((X\times V) \times W, (Y \times V) \times W) \\
}
\]
\end{mdefinition}

We say that a pair $(F, \sst)$ is a strong profunctor.
The diagrams that must commute here are similar to those for
a tensorial strength.

\begin{coder}
  The type class of strong profunctors is a simple extension of
  \ensuremath{\mathsf{Profunctor}}.

\begin{hscode}\SaveRestoreHook
\column{B}{@{}>{\hspre}l<{\hspost}@{}}%
\column{3}{@{}>{\hspre}l<{\hspost}@{}}%
\column{E}{@{}>{\hspre}l<{\hspost}@{}}%
\>[B]{}\mathkw{class}\;\mathsf{Profunctor}\;\Varid{p}\Rightarrow \mathsf{StrongProfunctor}\;\Varid{p}\;\mathkw{where}{}\<[E]%
\\
\>[B]{}\hsindent{3}{}\<[3]%
\>[3]{}\FN{first}\mathbin{::}\Varid{p}\;\Varid{x}\;\Varid{y}\to \Varid{p}\;(\Varid{x},\Varid{z})\;(\Varid{y},\Varid{z}){}\<[E]%
\ColumnHook
\end{hscode}\resethooks
Instances of the \ensuremath{\mathsf{StrongProfunctor}} class are subject to the following laws.
\begin{align*}
\ensuremath{\FN{dimap}\;\FN{id}\;\pi_1\;(\FN{first}\;\Varid{a})} &= \ensuremath{\FN{dimap}\;\pi_1\;\FN{id}\;\Varid{a}} \\
\ensuremath{\FN{first}\;(\FN{first}\;\Varid{a})} &= \ensuremath{\FN{dimap}\;\alpha \;\alpha^{-1}\;(\FN{first}\;\Varid{a})} \\
\ensuremath{\FN{dimap}\;(\FN{id}\;\times\;\Varid{f})\;\FN{id}\;(\FN{first}\;\Varid{a})} &= \ensuremath{\FN{dimap}\;\FN{id}\;(\FN{id}\;\times\;\Varid{f})\;(\FN{first}\;\Varid{a})}
\end{align*}
The first two laws correspond to the two diagrams above, while the
third one corresponds to dinaturality of \ensuremath{\FN{first}} in the \ensuremath{\Varid{z}} variable.
\end{coder}
In contrast to strong functors on $\Set$, the strength of a
profunctor may not exist, and even if it exists, it may not be unique.
\begin{coder}
  As an example of strengths not being unique, consider the following
  profunctor:
\begin{hscode}\SaveRestoreHook
\column{B}{@{}>{\hspre}l<{\hspost}@{}}%
\column{3}{@{}>{\hspre}l<{\hspost}@{}}%
\column{8}{@{}>{\hspre}l<{\hspost}@{}}%
\column{15}{@{}>{\hspre}l<{\hspost}@{}}%
\column{E}{@{}>{\hspre}l<{\hspost}@{}}%
\>[B]{}\mathkw{data}\;\mathsf{Double}\;\Varid{x}\;\Varid{y}\mathrel{=}\mathsf{Double}\;((\Varid{x},\Varid{x})\to (\Varid{y},\Varid{y})){}\<[E]%
\\[\blanklineskip]%
\>[B]{}\mathkw{instance}\;\mathsf{Profunctor}\;\mathsf{Double}\;\mathkw{where}{}\<[E]%
\\
\>[B]{}\hsindent{3}{}\<[3]%
\>[3]{}\FN{dimap}\;\Varid{f}\;\Varid{g}\;(\mathsf{Double}\;\Varid{h})\mathrel{=}\mathsf{Double}\;(\FN{lift}\;\Varid{g}\circ\Varid{h}\circ\FN{lift}\;\Varid{f}){}\<[E]%
\\
\>[3]{}\hsindent{5}{}\<[8]%
\>[8]{}\mathkw{where}\;{}\<[15]%
\>[15]{}\FN{lift}\mathbin{::}(\Varid{a}\to \Varid{b})\to (\Varid{a},\Varid{a})\to (\Varid{b},\Varid{b}){}\<[E]%
\\
\>[15]{}\FN{lift}\;\Varid{f}\;(\Varid{a},\Varid{a'})\mathrel{=}(\Varid{f}\;\Varid{a},\Varid{f}\;\Varid{a'}){}\<[E]%
\ColumnHook
\end{hscode}\resethooks
there exist two possible instances satisfying the strength axioms.
\begin{hscode}\SaveRestoreHook
\column{B}{@{}>{\hspre}l<{\hspost}@{}}%
\column{3}{@{}>{\hspre}l<{\hspost}@{}}%
\column{21}{@{}>{\hspre}l<{\hspost}@{}}%
\column{48}{@{}>{\hspre}l<{\hspost}@{}}%
\column{E}{@{}>{\hspre}l<{\hspost}@{}}%
\>[B]{}\mathkw{instance}\;\mathsf{StrongProfunctor}\;\mathsf{Double}\;\mathkw{where}{}\<[E]%
\\
\>[B]{}\hsindent{3}{}\<[3]%
\>[3]{}\FN{first}\;(\mathsf{Double}\;\Varid{f})\mathrel{=}\mathsf{Double}\;\Varid{g}{}\<[E]%
\\
\>[3]{}\hsindent{18}{}\<[21]%
\>[21]{}\mathkw{where}\;\Varid{g}\;((\Varid{x},\Varid{z}),(\Varid{x'},\Varid{z'})){}\<[48]%
\>[48]{}\mathrel{=}((\Varid{y},\Varid{z}),(\Varid{y'},\Varid{z'})){}\<[E]%
\\
\>[48]{}\mathkw{where}\;(\Varid{y},\Varid{y'})\mathrel{=}\Varid{f}\;(\Varid{x},\Varid{x'}){}\<[E]%
\ColumnHook
\end{hscode}\resethooks
\begin{hscode}\SaveRestoreHook
\column{B}{@{}>{\hspre}l<{\hspost}@{}}%
\column{3}{@{}>{\hspre}l<{\hspost}@{}}%
\column{21}{@{}>{\hspre}l<{\hspost}@{}}%
\column{48}{@{}>{\hspre}l<{\hspost}@{}}%
\column{E}{@{}>{\hspre}l<{\hspost}@{}}%
\>[B]{}\mathkw{instance}\;\mathsf{StrongProfunctor}\;\mathsf{Double}\;\mathkw{where}{}\<[E]%
\\
\>[B]{}\hsindent{3}{}\<[3]%
\>[3]{}\FN{first}\;(\mathsf{Double}\;\Varid{f})\mathrel{=}\mathsf{Double}\;\Varid{g}{}\<[E]%
\\
\>[3]{}\hsindent{18}{}\<[21]%
\>[21]{}\mathkw{where}\;\Varid{g}\;((\Varid{x},\Varid{z}),(\Varid{x'},\Varid{z'})){}\<[48]%
\>[48]{}\mathrel{=}((\Varid{y},\Varid{z}),(\Varid{y'},\Varid{z})){}\<[E]%
\\
\>[48]{}\mathkw{where}\;(\Varid{y},\Varid{y'})\mathrel{=}\Varid{f}\;(\Varid{x},\Varid{x'}){}\<[E]%
\ColumnHook
\end{hscode}\resethooks
Therefore, the profunctor \ensuremath{\mathsf{Double}} does not have a unique strength.
\end{coder}

Given two strong profunctors $(F, \sst^F)$, $(G, \sst^G)$, a \emph{strong
  natural transformation} is a natural transformation $\alpha : F \to
G$ that is compatible with the strengths:
\[
\xymatrix@C+=20mm{
F(X,Y) \ar[r]^-{\sst^F} \ar[d]_{\alpha} & F(X\times Z, Y\times Z) \ar[d]^{\alpha} \\
G(X,Y) \ar[r]_-{\sst^G} & G(X\times Z, Y\times Z)
}
\]

Following the approach to strong monads of Moggi~\shortcite{Moggi95},
we work with the category $\left[\C^{\op} \times \C,
  \Set\right]_{\str}$ of strong profunctors.

\begin{mdefinition}
  The category $\left[\C^{\op} \times \C, \Set\right]_{\str}$ consists
  of pairs $(F, \sst)$ as objects, where $F$ is a profunctor and $\sst$ is
  a strength for it, and strong natural transformations as morphisms.
\end{mdefinition}

Even when the strength for a functor is not unique, we usually write
$(F, \sst^F)$. Here the superscript $F$ in $\sst^F$ is just syntax to
distinguish between various strengths for different profunctors, but
it does not mean that $\sst^F$ is \emph{the} strength for $F$.

The monoidal structure of $\Pro$ can be used for strong
profunctors. Given two strong profunctors $(A,\sst^A)$ and $(B,\sst^B)$, a
family of morphisms $\sst^{A\oX B}_Z = \left[ \Lambda P . \iota_P \circ
  (\sst^A_Z \times \sst^B_Z) \right]$ is defined. It is easy to see that
such family is indeed a strength for the profunctor $A \oX B$. The
monoidal category of strong profunctors with tensor defined this way
is denoted by $\SPro$.

A monoid in $\SPro$ amounts to the same data that we had in the case
of $\Pro$.  This time, however, the morphisms $m$ and $e$ (being
morphisms of $\SPro$) must be compatible with the strength as well.
\begin{coder}
  Arrows can be implemented as strong profunctors which are pre-arrows.
\begin{hscode}\SaveRestoreHook
\column{B}{@{}>{\hspre}l<{\hspost}@{}}%
\column{E}{@{}>{\hspre}l<{\hspost}@{}}%
\>[B]{}\mathkw{class}\;(\mathsf{StrongProfunctor}\;\Varid{a},\mathsf{PreArrow}\;\Varid{a})\Rightarrow \mathsf{Arrow}\;\Varid{a}{}\<[E]%
\ColumnHook
\end{hscode}\resethooks

Instances of \ensuremath{\mathsf{Arrow}} are empty, but the programmer should check the
compatibility of the unit and multiplication of the pre-arrow with the
strength:
\begin{align*}
\ensuremath{\FN{first}\;(\FN{arr}\;\Varid{f})} &= \ensuremath{\FN{arr}\;(\Varid{f}\;\times\;\FN{id})} \\
\ensuremath{\FN{first}\;(\Varid{a}\ensuremath{\ggg}\Varid{b})} &= \ensuremath{\FN{first}\;\Varid{a}\ensuremath{\ggg}\FN{first}\;\Varid{b}}
\end{align*}
These two laws, together with the laws for profunctors, pre-arrows and
strength, constitute the arrows laws proposed by
Paterson~\shortcite{PatersonRA:fop}.
\end{coder}

\subsection{Exponential for Arrows}

Unfortunately, we have not managed to find an exponential for arrows.
Part of the difficulty in finding one seems to stem from the fact that
strengths for profunctors may not exist, and even if they do, they may
not be unique.  For example, given two strong profunctors $A$ and $B$,
the profunctor $B^A$ defined in
Section~\ref{sec:exponential_prearrows}, does not seem to have a
strength.

However, as shown next, it is possible to co-freely add a strength to
a pre-arrow, and use that to obtain a representation for arrows.

\subsection{Adding a Strength to Pre-Arrows}
\label{sec:tambara}

We mitigate the failure to find an exponential in $\SPro$ by building
on the success in $\Pro$. Concretely, we investigate how to add a
strength to profunctors in $\Pro$, and use this to complete the
constructions in $\Pro$ to make them work in $\SPro$.

There is an obvious monoidal functor from the monoidal category of
strong profunctors $\SPro$ into the monoidal category of profunctors
$\Pro$ that forgets the additional structure. More precisely, the functor
$U : \SPro \to \Pro$ forgets the strength.
  $$U (A,\sst^A) = A$$

Interestingly, this functor has a right adjoint. That is, there is a functor $T$
such that we have a natural isomorphism.
\begin{equation}
\label{eq:tambara}
 \phi : \Pro(U (A, \sst^A), B) \cong \SPro((A, \sst^A), T B)   
\end{equation}

The monoidal functor $T :\Pro \to\SPro$ is given by $T B = (T_B , \sst)$, with
its components defined by
\[
\begin{aligned}
T_B(X,Y) &= \int_Z B(X \times Z, Y \times Z) \\
\sst_Z &= \langle \Lambda V.\, B(\alpha, \alpha^{-1}) \circ \omega_{Z \times V} \rangle
\end{aligned}
\]

The adjunction $U \dashv T$ tells us that $T$ completes a profunctors
by (co)freely adding a strength.
Pastro and Street, when working on Tambara modules~\cite{pastro2008},
introduced an endofunctor on the category of profunctors which adds a
structure similar to what we call a strength. The functor $T$ is
based on that endofunctor and hence we call it the \emph{Tambara functor}.

\begin{coder}
The Tambara functor may be implemented as follows.

\begin{hscode}\SaveRestoreHook
\column{B}{@{}>{\hspre}l<{\hspost}@{}}%
\column{3}{@{}>{\hspre}l<{\hspost}@{}}%
\column{20}{@{}>{\hspre}l<{\hspost}@{}}%
\column{27}{@{}>{\hspre}l<{\hspost}@{}}%
\column{35}{@{}>{\hspre}l<{\hspost}@{}}%
\column{47}{@{}>{\hspre}l<{\hspost}@{}}%
\column{E}{@{}>{\hspre}l<{\hspost}@{}}%
\>[B]{}\mathkw{data}\;\mathsf{Tambara}\;\Varid{a}\;\Varid{x}\;\Varid{y}\mathrel{=}\mathsf{Tambara}\;(\forall \Varid{z}\hsforall .\ \Varid{a}\;(\Varid{x},\Varid{z})\;(\Varid{y},\Varid{z})){}\<[E]%
\\[\blanklineskip]%
\>[B]{}\mathkw{instance}\;\mathsf{Profunctor}\;\Varid{a}\Rightarrow \mathsf{Profunctor}\;(\mathsf{Tambara}\;\Varid{a})\;\mathkw{where}{}\<[E]%
\\
\>[B]{}\hsindent{3}{}\<[3]%
\>[3]{}\FN{dimap}\;\Varid{f}\;\Varid{g}\;(\mathsf{Tambara}\;\Varid{x})\mathrel{=}\mathsf{Tambara}\;(\FN{dimap}\;(\FN{lift}\;\Varid{f})\;(\FN{lift}\;\Varid{g})\;\Varid{x}){}\<[E]%
\\
\>[3]{}\hsindent{24}{}\<[27]%
\>[27]{}\mathkw{where}\;\FN{lift}\;\Varid{f}\;(\Varid{a},\Varid{b})\mathrel{=}(\Varid{f}\;\Varid{a},\Varid{b}){}\<[E]%
\\[\blanklineskip]%
\>[B]{}\mathkw{instance}\;\mathsf{Profunctor}\;\Varid{a}\Rightarrow \mathsf{StrongProfunctor}\;(\mathsf{Tambara}\;\Varid{a})\;\mathkw{where}{}\<[E]%
\\
\>[B]{}\hsindent{3}{}\<[3]%
\>[3]{}\FN{first}\;(\mathsf{Tambara}\;\Varid{x})\mathrel{=}\mathsf{Tambara}\;(\FN{dimap}\;\alpha \;\alpha^{-1}\;\Varid{x}){}\<[E]%
\\
\>[3]{}\hsindent{17}{}\<[20]%
\>[20]{}\mathkw{where}\;{}\<[27]%
\>[27]{}\alpha \;{}\<[35]%
\>[35]{}((\Varid{x},\Varid{y}),\Varid{z}){}\<[47]%
\>[47]{}\mathrel{=}(\Varid{x},(\Varid{y},\Varid{z})){}\<[E]%
\\
\>[27]{}\alpha^{-1}\;{}\<[35]%
\>[35]{}(\Varid{x},(\Varid{y},\Varid{z})){}\<[47]%
\>[47]{}\mathrel{=}((\Varid{x},\Varid{y}),\Varid{z}){}\<[E]%
\ColumnHook
\end{hscode}\resethooks
\vspace{-5mm}
\end{coder}

The isomorphism~\ref{eq:tambara} is witnessed by morphisms:
\[
\begin{aligned}
\left(\phi(\eta)_{(A,\sst^A),B}\right)_{X,Y} &= \langle \Lambda Z . \eta_{X\times Z, Y\times Z} \circ \sst^A_Z \rangle \\
\left(\phi^{-1}(\beta)_{(A,\sst^A),B}\right)_{X,Y} &= B({\pi_1}^{-1}, \pi_1) \circ \omega_1 \circ \beta_{X,Y}
\end{aligned}
\]%

\begin{coder}
  The components of the isomorphism~\ref{eq:tambara} are implemented
  in Haskell as follows.
\begin{hscode}\SaveRestoreHook
\column{B}{@{}>{\hspre}l<{\hspost}@{}}%
\column{12}{@{}>{\hspre}l<{\hspost}@{}}%
\column{19}{@{}>{\hspre}l<{\hspost}@{}}%
\column{E}{@{}>{\hspre}l<{\hspost}@{}}%
\>[B]{}\phi\mathbin{::}(\mathsf{StrongProfunctor}\;\Varid{a},\mathsf{Profunctor}\;\Varid{b})\Rightarrow (\Varid{a}\xrightarrow{\sbt\ \sbt}\Varid{b})\to (\Varid{a}\xrightarrow{\sbt\ \sbt}\mathsf{Tambara}\;\Varid{b}){}\<[E]%
\\
\>[B]{}\phi\;\Varid{f}\;\Varid{a}\mathrel{=}\mathsf{Tambara}\;(\Varid{f}\;(\FN{first}\;\Varid{a})){}\<[E]%
\\[\blanklineskip]%
\>[B]{}\phi^{-1}\mathbin{::}(\mathsf{StrongProfunctor}\;\Varid{a},\mathsf{Profunctor}\;\Varid{b})\Rightarrow (\Varid{a}\xrightarrow{\sbt\ \sbt}\mathsf{Tambara}\;\Varid{b})\to (\Varid{a}\xrightarrow{\sbt\ \sbt}\Varid{b}){}\<[E]%
\\
\>[B]{}\phi^{-1}\;\Varid{f}\;\Varid{a}\mathrel{=}\FN{dimap}\;\FN{fst}^{-1}\;\FN{fst}\;\Varid{b}{}\<[E]%
\\
\>[B]{}\hsindent{12}{}\<[12]%
\>[12]{}\mathkw{where}\;{}\<[19]%
\>[19]{}\mathsf{Tambara}\;\Varid{b}\mathrel{=}\Varid{f}\;\Varid{a}{}\<[E]%
\\
\>[19]{}\FN{fst}^{-1}\;\Varid{x}\mathrel{=}(\Varid{x},()){}\<[E]%
\ColumnHook
\end{hscode}\resethooks
\vspace{-5mm}
\end{coder}

Since we have an adjunction $U \dashv T$, we can form a comonad $UT :
\Pro \to \Pro$.  The counit of $UT$ is the counit of the adjunction:
$\epsilon = \phi^{-1}(\id : A \to A) = A({\pi_1}^{-1}, \pi_1) \circ \omega_1$, 
and its comultiplication is 
$\delta = U \phi(\id : UT A \to UT A) T 
        = \langle \Lambda Z .\, 
                     \langle 
                        \Lambda V .\, A(\alpha, \alpha^{-1}) \circ \omega_{Z \times V} 
                      \rangle 
          \rangle$.

\begin{mprop}
The category $\SPro$ is equivalent to the
(co)Eilenberg-Moore category for the comonad $UT$.  
\end{mprop}

\begin{mproof}
A coalgebra
for this comonad is an object $A$ from $\Pro$ together with a morphism
$\sigma : A \to UT A$ such that these diagrams commute:
\[
\xymatrix@C+=5mm{
  UT A \ar@2{-}[rd] \ar[r]^-{\sigma} & UT (UT A) \ar[d]^{\epsilon} \\
       & UT A \\
}
\qquad
\xymatrix@C+=5mm{
  A \ar[r]^-{\sigma} \ar[d]_-{\sigma} & UT A \ar[d]^{UT \sigma} \\
  UT A \ar[r]_-{\delta} & UT (UT A) \\
}
\]

\noindent The morphism $\sigma$ is a family of morphisms
\[ \sigma_{X,Y} : A(X,Y) \to \int_Z A(X \times Z, Y \times Z) \]%
natural in $X$ and $Y$. By the universal property of ends, the family
$\sigma_{X,Y}$ is equivalent to a family of morphisms
\[ \bar{\sigma}_{X,Y,Z} : A(X,Y) \to A(X \times Z, Y \times Z) \]%
natural in $X$ and $Y$\!, and dinatural in $Z$.
Such family has exactly the form of a strength. Using the laws for coalgebras of a comonad, one can prove that $\bar{\sigma}$ is
indeed a strength for $A$. 
\end{mproof}

\subsection{A Representation of Arrows}
Although we have not found the form of exponential objects in $\SPro$,
an alternative representation for monoids can be given with help of
the Tambara functor.

The idea is to take a monoid in $\SPro$ and forget the strength
structure using $U$. Then, use the Cayley representation for monoids
in $\Pro$, and finally apply the Tambara functor to obtain a new
strength on this monoid.
That is, given a monoid $(M,m,e)$ in $\SPro$, its representation is
$T(UM^{UM})$.  The functor $T$ is monoidal and therefore, as shown by
Theorem~\ref{thm:monoid}, it takes monoids in $\Pro$ to monoids in
$\SPro$.

More concretely, given a monoid $((A,{\sst}^A), m, e)$ in $\SPro$
(i.e. and arrow), we construct a morphism $\rep : (A, {\sst}^A) \to T
A^A$ as %
\[
\rep_{X,Y} = \langle \Lambda Z . \langle \Lambda D. \ladj{m \circ \iota_{Y\times Z} \circ ({\sst}^{A}_{Z} \times \id)} \rangle \rangle
\] %
This is a legit morphism in $\SPro$, i.e. it commutes with the
strengths of $A$ and $T A^A$.  It has a left inverse, 
 $\abs : U T A^A \to U (A, {\sst}^A)$ defined as
\[ \abs_{X,Y} =       A({\pi_1}^{-1}, \id) 
               \circ \ev 
               \circ (\id \times e) 
               \circ \langle \id, \ladj{\pi_1 \circ \pi_2} \circ ! \rangle 
               \circ \omega_{Y}
               \circ \omega_{1}
\]%
and therefore, $\rep$ is a monomorphism. 
This proves that $T A^A$ is a representation for $(A, {\sst}^A)$.

\begin{coder}
The representation is implemented in Haskell as follows.

\begin{hscode}\SaveRestoreHook
\column{B}{@{}>{\hspre}l<{\hspost}@{}}%
\column{3}{@{}>{\hspre}l<{\hspost}@{}}%
\column{23}{@{}>{\hspre}l<{\hspost}@{}}%
\column{E}{@{}>{\hspre}l<{\hspost}@{}}%
\>[B]{}\mathkw{data}\;\mathsf{Rep}\;\Varid{a}\;\Varid{x}\;\Varid{y}\mathrel{=}\mathsf{Rep}\;(\forall \Varid{z'}\hsforall \;\Varid{z}.\ \Varid{a}\;(\Varid{y},\Varid{z'})\;\Varid{z}\to \Varid{a}\;(\Varid{x},\Varid{z'})\;\Varid{z}){}\<[E]%
\\[\blanklineskip]%
\>[B]{}\mathkw{instance}\;\mathsf{Profunctor}\;\Varid{a}\Rightarrow \mathsf{Profunctor}\;(\mathsf{Rep}\;\Varid{a})\;\mathkw{where}{}\<[E]%
\\
\>[B]{}\hsindent{3}{}\<[3]%
\>[3]{}\FN{dimap}\;\Varid{f}\;\Varid{g}\;(\mathsf{Rep}\;\Varid{x})\mathrel{=}\mathsf{Rep}\;(\lambda \Varid{y}\to \FN{dimap}\;(\FN{lift}\;\Varid{f})\;\FN{id}\;(\Varid{x}\;(\FN{dimap}\;(\FN{lift}\;\Varid{g})\;\FN{id}\;\Varid{y}))){}\<[E]%
\\
\>[3]{}\hsindent{20}{}\<[23]%
\>[23]{}\mathkw{where}\;\FN{lift}\;\Varid{f}\;(\Varid{a},\Varid{b})\mathrel{=}(\Varid{f}\;\Varid{a},\Varid{b}){}\<[E]%
\ColumnHook
\end{hscode}\resethooks

The representation takes any profunctor into an arrow.
\begin{hscode}\SaveRestoreHook
\column{B}{@{}>{\hspre}l<{\hspost}@{}}%
\column{3}{@{}>{\hspre}l<{\hspost}@{}}%
\column{19}{@{}>{\hspre}l<{\hspost}@{}}%
\column{26}{@{}>{\hspre}l<{\hspost}@{}}%
\column{E}{@{}>{\hspre}l<{\hspost}@{}}%
\>[B]{}\mathkw{instance}\;\mathsf{Profunctor}\;\Varid{a}\Rightarrow \mathsf{PreArrow}\;(\mathsf{Rep}\;\Varid{a})\;\mathkw{where}{}\<[E]%
\\
\>[B]{}\hsindent{3}{}\<[3]%
\>[3]{}\FN{arr}\;\Varid{f}\mathrel{=}\mathsf{Rep}\;(\FN{dimap}\;(\FN{lift}\;\Varid{f})\;\FN{id})\;\mathkw{where}\;\FN{lift}\;\Varid{f}\;(\Varid{a},\Varid{b})\mathrel{=}(\Varid{f}\;\Varid{a},\Varid{b}){}\<[E]%
\\
\>[B]{}\hsindent{3}{}\<[3]%
\>[3]{}\mathsf{Rep}\;\Varid{x}\ensuremath{\ggg}\mathsf{Rep}\;\Varid{y}\mathrel{=}\mathsf{Rep}\;(\lambda \Varid{v}\to \Varid{x}\;(\Varid{y}\;\Varid{v})){}\<[E]%
\\[\blanklineskip]%
\>[B]{}\mathkw{instance}\;\mathsf{Profunctor}\;\Varid{a}\Rightarrow \mathsf{StrongProfunctor}\;(\mathsf{Rep}\;\Varid{a})\;\mathkw{where}{}\<[E]%
\\
\>[B]{}\hsindent{3}{}\<[3]%
\>[3]{}\FN{first}\;(\mathsf{Rep}\;\Varid{x})\mathrel{=}\mathsf{Rep}\;(\lambda \Varid{z}\to \FN{dimap}\;\alpha \;\FN{id}\;(\Varid{x}\;(\FN{dimap}\;\alpha^{-1}\;\FN{id}\;\Varid{z}))){}\<[E]%
\\
\>[3]{}\hsindent{16}{}\<[19]%
\>[19]{}\mathkw{where}\;{}\<[26]%
\>[26]{}\alpha \;((\Varid{x},\Varid{y}),\Varid{z})\mathrel{=}(\Varid{x},(\Varid{y},\Varid{z})){}\<[E]%
\\
\>[26]{}\alpha^{-1}\;(\Varid{x},(\Varid{y},\Varid{z}))\mathrel{=}((\Varid{x},\Varid{y}),\Varid{z}){}\<[E]%
\ColumnHook
\end{hscode}\resethooks

Since we verified that the strength is compatible with the pre-arrow
structure, we may declare the \ensuremath{\mathsf{Arrow}} instance.
\begin{hscode}\SaveRestoreHook
\column{B}{@{}>{\hspre}l<{\hspost}@{}}%
\column{E}{@{}>{\hspre}l<{\hspost}@{}}%
\>[B]{}\mathkw{instance}\;\mathsf{Profunctor}\;\Varid{a}\Rightarrow \mathsf{Arrow}\;(\mathsf{Rep}\;\Varid{a}){}\<[E]%
\ColumnHook
\end{hscode}\resethooks

Any arrow \ensuremath{\Varid{a}} can be represented by \ensuremath{\mathsf{Rep}\;\Varid{a}}. Moreover, \ensuremath{\abs\circ\rep\mathrel{=}\FN{id}}. 
\begin{hscode}\SaveRestoreHook
\column{B}{@{}>{\hspre}l<{\hspost}@{}}%
\column{13}{@{}>{\hspre}l<{\hspost}@{}}%
\column{33}{@{}>{\hspre}l<{\hspost}@{}}%
\column{E}{@{}>{\hspre}l<{\hspost}@{}}%
\>[B]{}\rep\mathbin{::}\mathsf{Arrow}\;\Varid{a}\Rightarrow \Varid{a}\;\Varid{x}\;\Varid{y}\to \mathsf{Rep}\;\Varid{a}\;\Varid{x}\;\Varid{y}{}\<[E]%
\\
\>[B]{}\rep\;\Varid{x}\mathrel{=}\mathsf{Rep}\;(\lambda \Varid{z}\to \FN{first}\;\Varid{x}\ensuremath{\ggg}\Varid{z}){}\<[E]%
\\[\blanklineskip]%
\>[B]{}\abs\mathbin{::}\mathsf{Arrow}\;\Varid{a}\Rightarrow \mathsf{Rep}\;\Varid{a}\;\Varid{x}\;\Varid{y}\to \Varid{a}\;\Varid{x}\;\Varid{y}{}\<[E]%
\\
\>[B]{}\abs\;(\mathsf{Rep}\;\Varid{x})\mathrel{=}\FN{arr}\;\FN{fst}^{-1}\ensuremath{\ggg}\Varid{x}\;(\FN{arr}\;\FN{fst}){}\<[E]%
\\
\>[B]{}\hsindent{13}{}\<[13]%
\>[13]{}\mathkw{where}\;\FN{fst}^{-1}\;\Varid{y}{}\<[33]%
\>[33]{}\mathrel{=}(\Varid{y},()){}\<[E]%
\ColumnHook
\end{hscode}\resethooks
\vspace{-5mm}
\end{coder}

\subsection{Free Arrows}

Having failed to find an exponential in $\SPro$, we cannot apply
Proposition~\ref{prop:free} to obtain the free monoid in $\SPro$ and
therefore we fall back to finding it directly. Fortunately, we do not
need to search much as the free monoid on $\Pro$ is equipped with an obvious
strength whenever it is built over a strong profunctor, and indeed one
can verify that the obtained monoid is the free monoid in $\SPro$.

\begin{coder}
  The free pre-arrow can be equipped with a strength when defined over
  a strong profunctor.
\begin{hscode}\SaveRestoreHook
\column{B}{@{}>{\hspre}l<{\hspost}@{}}%
\column{3}{@{}>{\hspre}l<{\hspost}@{}}%
\column{21}{@{}>{\hspre}l<{\hspost}@{}}%
\column{E}{@{}>{\hspre}l<{\hspost}@{}}%
\>[B]{}\mathkw{instance}\;\mathsf{StrongProfunctor}\;\Varid{a}\Rightarrow \mathsf{StrongProfunctor}\;(\FN{Free}_{\oX}\;\Varid{a})\;\mathkw{where}{}\<[E]%
\\
\>[B]{}\hsindent{3}{}\<[3]%
\>[3]{}\FN{first}\;(\mathsf{Hom}\;\Varid{f}){}\<[21]%
\>[21]{}\mathrel{=}\mathsf{Hom}\;(\lambda (\Varid{x},\Varid{z})\to (\Varid{f}\;\Varid{x},\Varid{z})){}\<[E]%
\\
\>[B]{}\hsindent{3}{}\<[3]%
\>[3]{}\FN{first}\;(\mathsf{Comp}\;\Varid{x}\;\Varid{y}){}\<[21]%
\>[21]{}\mathrel{=}\mathsf{Comp}\;(\FN{first}\;\Varid{x})\;(\FN{first}\;\Varid{y}){}\<[E]%
\ColumnHook
\end{hscode}\resethooks
Since the unit and multiplication of the free arrow are compatible
with the strength, we can declare the \ensuremath{\mathsf{Arrow}} instance without guilt.

\begin{hscode}\SaveRestoreHook
\column{B}{@{}>{\hspre}l<{\hspost}@{}}%
\column{E}{@{}>{\hspre}l<{\hspost}@{}}%
\>[B]{}\mathkw{instance}\;\mathsf{StrongProfunctor}\;\Varid{a}\Rightarrow \mathsf{Arrow}\;(\FN{Free}_{\oX}\;\Varid{a}){}\<[E]%
\ColumnHook
\end{hscode}\resethooks
The insertion of generators and the universal morphism are the same as
the ones from pre-arrows. The only difference is that now we require
\ensuremath{\mathsf{StrongProfunctor}}s instead of plain \ensuremath{\mathsf{Profunctor}}s. 
\begin{hscode}\SaveRestoreHook
\column{B}{@{}>{\hspre}l<{\hspost}@{}}%
\column{20}{@{}>{\hspre}l<{\hspost}@{}}%
\column{E}{@{}>{\hspre}l<{\hspost}@{}}%
\>[B]{}\ins\mathbin{::}\mathsf{StrongProfunctor}\;\Varid{a}\Rightarrow \Varid{a}\xrightarrow{\sbt\ \sbt}\FN{Free}_{\oX}\;\Varid{a}{}\<[E]%
\\
\>[B]{}\ins\;\Varid{x}\mathrel{=}\mathsf{Comp}\;\Varid{x}\;(\FN{arr}\;\FN{id}){}\<[E]%
\\[\blanklineskip]%
\>[B]{}\FN{free}\mathbin{::}(\mathsf{StrongProfunctor}\;\Varid{a},\mathsf{Arrow}\;\Varid{b})\Rightarrow (\Varid{a}\xrightarrow{\sbt\ \sbt}\Varid{b})\to (\FN{Free}_{\oX}\;\Varid{a}\xrightarrow{\sbt\ \sbt}\Varid{b}){}\<[E]%
\\
\>[B]{}\FN{free}\;\Varid{f}\;(\mathsf{Hom}\;\Varid{g}){}\<[20]%
\>[20]{}\mathrel{=}\FN{arr}\;\Varid{g}{}\<[E]%
\\
\>[B]{}\FN{free}\;\Varid{f}\;(\mathsf{Comp}\;\Varid{x}\;\Varid{y}){}\<[20]%
\>[20]{}\mathrel{=}\Varid{f}\;\Varid{x}\ensuremath{\ggg}\FN{free}\;\Varid{f}\;\Varid{y}{}\<[E]%
\ColumnHook
\end{hscode}\resethooks

Here, we would really like the type \ensuremath{(\Varid{a}\xrightarrow{\sbt\ \sbt}\Varid{b})} to represent
strength preserving morphisms between strong profunctors. Therefore,
\ensuremath{\FN{free}\;\Varid{f}} will preserve the strength only when \ensuremath{\Varid{f}} does.
\end{coder}

\begin{mremark}
We now have adjunctions
\[
\xymatrix{  \Pro 
           \ar@/_2.5ex/[r]_{T}
            \ar@{}[r]|-{\perp} 
          & \SPro 
           \ar@/_2.5ex/[l]_{U}
           \ar@/^3ex/[r]^{-^*}
            \ar@{}[r]|-{\perp} 
          & \Monoid{\SPro}.
           \ar@/^3ex/[l]^{U}
}
\]%
However, they do not compose. Free arrows
are generated freely from $\SPro$ but the strength is generated
co-freely from $\Pro$. This provides an explanation for the
difficulty in defining a free arrow over an arbitrary (weak) profunctor.
\end{mremark}




%

\section{On Functors Between Monoidal Categories}
\label{sec:functors}
Monads, applicatives and arrows have been introduced as monoids in
monoidal categories. Now we ask what is the relation between these
monoidal categories. It is well-known that from a monad it can be
derived both an applicative functor and an arrow. In this section we
explain these and other derivations from the point of view of
monoidal categories.

For example, in order to obtain a pre-arrow from a monad, we are
interested in creating a monoid in $\Pro$, given a monoid in $\Endo$.
Instead of trying to make up a monoid in $\Pro$ directly, we will
define a monoidal functor between the underlying monoidal categories
(in this case $\Endo$ and $\Pro$), and then use the following theorem
to obtain a functor between the corresponding monoids.

\begin{mtheorem}
\label{thm:monoid}
Let $(F, \phi, \eta) : \C \to \D$ be a lax monoidal functor. If $(M, m,
e)$ is a monoid in $\C$, then $(F M, F m \circ \phi, F e \circ \eta)$
is a monoid in $\D$.
\end{mtheorem}

\extended{
\begin{mproof}
\begin{align*}
   & F m \circ \phi \circ (\id \oX (F e \circ \eta))\\[\nl]
=  &  \comment{tensor} \\[\nl]
   & F m \circ \phi \circ (\id \oX F e) \circ (\id \circ \eta) \\[\nl]
=  &  \comment{naturality} \\[\nl]
   & F m \circ F (\id \oX e) \circ \phi \circ (\id \circ \eta) \\[\nl]
=  &  \comment{$F$ functor} \\[\nl]
   & F (m \circ (\id \oX e)) \circ \phi \circ (\id \circ \eta) \\[\nl]
=  &  \comment{monoid} \\[\nl]
   & F \rho_\C \circ \phi \circ (\id \circ \eta) \\[\nl]
=  &  \comment{monoidal functor} \\[\nl]
   & \rho_\D
 \end{align*}
\[
\begin{align*}
   & F m \circ \phi \circ (\id \oX (F m \circ \phi))\\[\nl]
=  &  \comment{tensor} \\[\nl]
   & F m \circ \phi \circ (\id \oX F m) \circ (\id \circ \phi) \\[\nl]
=  &  \comment{natural} \\[\nl]
   & F m \circ F (\id \oX m) \circ \phi \circ (\id \circ \phi) \\[\nl]
=  &  \comment{$F$ functor} \\[\nl]
   & F (m \circ (\id \oX m)) \circ \phi \circ (\id \circ \phi) \\[\nl]
=  &  \comment{monoid} \\[\nl]
   & F (m \circ (m \oX \id) \circ \alpha) \circ \phi \circ (\id \circ \phi) \\[\nl]
=  &  \comment{$F$ functor} \\[\nl]
   & F m \circ F (m \oX \id) \circ F \alpha \circ \phi \circ (\id \circ \phi) \\[\nl]
=  &  \comment{monoidal functor} \\[\nl]
   & F m \circ F (m \oX \id) \circ \phi \circ (\phi \circ \id) \circ \alpha \\[\nl]
=  &  \comment{natural} \\[\nl]
   & F m \circ \phi \circ (F m \oX \id) \circ (\phi \circ \id) \circ \alpha \\[\nl]
=  &  \comment{tensor} \\[\nl]
   & F m \circ \phi \circ ((F m \circ \phi) \oX \id) \circ \alpha
\end{align*}
\]
\end{mproof}
}
\noindent The above construction extends to a functor, and therefore we can
induce functors between monoids by way of lax monoidal functors between their
underlying monoidal categories.


\subsection{The Cayley Functor}
Applicative functors can be used to create arrows, here we present a
monoidal functor that gives rise to such construction.
We consider the \emph{Cayley functor}~\cite{pastro2008}
\[ C : \EndD \to \SPro \]%
defined by
\[ C(F)(X,Y) = F(Y^X) \]%
Despite its name, this functor bears no direct relation with the Cayley representation.

\begin{mprop}
The Cayley functor is monoidal from $\EndD$ to $\Pro$.
\end{mprop}

Not only this functor is monoidal but also, for each
$C(F)$, we can define a strength. This extends $C$ into
a monoidal functor from $\EndD$ to $\SPro$.

\begin{coder}
The resulting construction is the static arrow over \ensuremath{(\to )}, augmented
with the original applicative~\cite{mcbride08:applicative-programming}.
\begin{hscode}\SaveRestoreHook
\column{B}{@{}>{\hspre}l<{\hspost}@{}}%
\column{E}{@{}>{\hspre}l<{\hspost}@{}}%
\>[B]{}\mathkw{data}\;\mathsf{Cayley}\;\Varid{f}\;\Varid{x}\;\Varid{y}\mathrel{=}\mathsf{Cayley}\;(\Varid{f}\;(\Varid{x}\to \Varid{y})){}\<[E]%
\ColumnHook
\end{hscode}\resethooks

For every applicative functor, the Cayley functor constructs an arrow. 
\begin{hscode}\SaveRestoreHook
\column{B}{@{}>{\hspre}l<{\hspost}@{}}%
\column{3}{@{}>{\hspre}l<{\hspost}@{}}%
\column{30}{@{}>{\hspre}l<{\hspost}@{}}%
\column{E}{@{}>{\hspre}l<{\hspost}@{}}%
\>[B]{}\mathkw{instance}\;\mathsf{Applicative}\;\Varid{f}\Rightarrow \mathsf{PreArrow}\;(\mathsf{Cayley}\;\Varid{f})\;\mathkw{where}{}\<[E]%
\\
\>[B]{}\hsindent{3}{}\<[3]%
\>[3]{}\FN{arr}\;\Varid{f}{}\<[30]%
\>[30]{}\mathrel{=}\mathsf{Cayley}\;(\FN{pure}\;\Varid{f}){}\<[E]%
\\
\>[B]{}\hsindent{3}{}\<[3]%
\>[3]{}(\mathsf{Cayley}\;\Varid{x})\ensuremath{\ggg}(\mathsf{Cayley}\;\Varid{y}){}\<[30]%
\>[30]{}\mathrel{=}\mathsf{Cayley}\;(\FN{pure}\;(\circ)\circledast\Varid{y}\circledast\Varid{x}){}\<[E]%
\\[\blanklineskip]%
\>[B]{}\mathkw{instance}\;\mathsf{Applicative}\;\Varid{f}\Rightarrow \mathsf{StrongProfunctor}\;(\mathsf{Cayley}\;\Varid{f})\;\mathkw{where}{}\<[E]%
\\
\>[B]{}\hsindent{3}{}\<[3]%
\>[3]{}\FN{first}\;(\mathsf{Cayley}\;\Varid{x}){}\<[30]%
\>[30]{}\mathrel{=}\mathsf{Cayley}\;(\FN{pure}\;(\lambda \Varid{f}\to \lambda (\Varid{b},\Varid{d})\to (\Varid{f}\;\Varid{b},\Varid{d}))\circledast\Varid{x}){}\<[E]%
\\[\blanklineskip]%
\>[B]{}\mathkw{instance}\;\mathsf{Applicative}\;\Varid{f}\Rightarrow \mathsf{Arrow}\;(\mathsf{Cayley}\;\Varid{f}){}\<[E]%
\ColumnHook
\end{hscode}\resethooks
\vspace{-5mm}
\end{coder}

\subsection{The Kleisli Functor}
The well-known Kleisli category of a monad gives rise to
a monoidal functor from monads to arrows.
We consider the functor
\[ K : \EndC \to \SPro \]%
defined by
\[ K(F)(X,Y) = (F(Y))^X \]%
\begin{coder}
  The implementation of the Kleisli functor is as follows.
\begin{hscode}\SaveRestoreHook
\column{B}{@{}>{\hspre}l<{\hspost}@{}}%
\column{3}{@{}>{\hspre}l<{\hspost}@{}}%
\column{32}{@{}>{\hspre}l<{\hspost}@{}}%
\column{E}{@{}>{\hspre}l<{\hspost}@{}}%
\>[B]{}\mathkw{data}\;\mathsf{Kleisli}\;\Varid{f}\;\Varid{x}\;\Varid{y}\mathrel{=}\mathsf{Kleisli}\;(\Varid{x}\to \Varid{f}\;\Varid{y}){}\<[E]%
\\[\blanklineskip]%
\>[B]{}\mathkw{instance}\;\mathsf{Monad}\;\Varid{f}\Rightarrow \mathsf{PreArrow}\;(\mathsf{Kleisli}\;\Varid{f})\;\mathkw{where}{}\<[E]%
\\
\>[B]{}\hsindent{3}{}\<[3]%
\>[3]{}\FN{arr}\;\Varid{f}{}\<[32]%
\>[32]{}\mathrel{=}\mathsf{Kleisli}\;(\lambda \Varid{x}\to \FN{return}\;(\Varid{f}\;\Varid{x})){}\<[E]%
\\
\>[B]{}\hsindent{3}{}\<[3]%
\>[3]{}(\mathsf{Kleisli}\;\Varid{f})\ensuremath{\ggg}(\mathsf{Kleisli}\;\Varid{g}){}\<[32]%
\>[32]{}\mathrel{=}\mathsf{Kleisli}\;(\lambda \Varid{x}\to \Varid{f}\;\Varid{x}\bind \Varid{g}){}\<[E]%
\\[\blanklineskip]%
\>[B]{}\mathkw{instance}\;\mathsf{Monad}\;\Varid{f}\Rightarrow \mathsf{StrongProfunctor}\;(\mathsf{Kleisli}\;\Varid{f})\;\mathkw{where}{}\<[E]%
\\
\>[B]{}\hsindent{3}{}\<[3]%
\>[3]{}\FN{first}\;(\mathsf{Kleisli}\;\Varid{f}){}\<[32]%
\>[32]{}\mathrel{=}\mathsf{Kleisli}\;(\lambda (\Varid{b},\Varid{d})\to \Varid{f}\;\Varid{b}\bind \lambda \Varid{c}\to \FN{return}\;(\Varid{c},\Varid{d})){}\<[E]%
\\[\blanklineskip]%
\>[B]{}\mathkw{instance}\;\mathsf{Monad}\;\Varid{f}\Rightarrow \mathsf{Arrow}\;(\mathsf{Kleisli}\;\Varid{f}){}\<[E]%
\ColumnHook
\end{hscode}\resethooks
\vspace{-5mm}
\end{coder}

\subsection{The Identity Functor}
\label{sec:identity}

The identity endofunctor on $[\Set,\Set]$ can be given a monoidal
compatibility morphisms $\eta$ and $\phi$ such that it becomes a
lax monoidal functor from $\EndC$ to $\EndD$. The $\eta$ morphism is
the identity on the identity functor. The morphism $\phi_{F,G}
: F \dayt G \to F \circ G$ is given by:
\begin{align*}
  (F\dayt G)(A) = F C \times G D \times A^{(C\times D)} &\stackrel{\funst}{\verylongrightarrow} F (C \times (G D \times A ^{(C\times D)})) \\[\nl]
  &\stackrel{F \funst}{\verylongrightarrow} F (G (C \times D \times A^{(C\times D)})) \\[\nl]
  &\stackrel{F (G\, \ev)}{\verylongrightarrow} F (G A)
\end{align*}

Hence, we obtain a lax monoidal functor $\hat\Id : \EndC \to \EndD$.

\begin{coder}
  By applying Theorem~\ref{thm:monoid} to 
  $\hat\Id$, we obtain the well-known result that every monad is an
  applicative functor.

\begin{hscode}\SaveRestoreHook
\column{B}{@{}>{\hspre}l<{\hspost}@{}}%
\column{3}{@{}>{\hspre}l<{\hspost}@{}}%
\column{12}{@{}>{\hspre}l<{\hspost}@{}}%
\column{E}{@{}>{\hspre}l<{\hspost}@{}}%
\>[B]{}\mathkw{instance}\;\mathsf{Monad}\;\Varid{f}\Rightarrow \mathsf{Applicative}\;\Varid{f}\;\mathkw{where}{}\<[E]%
\\
\>[B]{}\hsindent{3}{}\<[3]%
\>[3]{}\FN{pure}{}\<[12]%
\>[12]{}\mathrel{=}\FN{return}{}\<[E]%
\\
\>[B]{}\hsindent{3}{}\<[3]%
\>[3]{}\Varid{f}\circledast\Varid{x}{}\<[12]%
\>[12]{}\mathrel{=}\Varid{f}\bind (\lambda \Varid{g}\to \Varid{x}\bind \FN{return}\circ\Varid{g}){}\<[E]%
\ColumnHook
\end{hscode}\resethooks
\vspace{-5mm}
\end{coder}

\subsection{The Reversed Monoid}
For every monoidal category $\C_{\oX} = (\C, \oX, \oI, \alpha, \lambda, \rho)$,
the opposite monoidal category $\C_{{\oX}^{\op}}$ can be defined,
with the monoidal operator $A {\oX}^{\op} B = B \oX A$. Given
a monoid in $\C_{\oX}$, a monoid in $\C_{{\oX}^{\op}}$ can be defined.
\begin{mtheorem}
\label{thm:monoidop}
If $(M, m, e)$ is a monoid in $\C_{\oX}$, then $(M, m, e)$ is
a monoid in $\C_{{\oX}^{\op}}$.
\end{mtheorem}
In the case where the monoidal structure is symmetric, there is an
isomorphism between $A {\oX}^{\op} B$ and $A \oX B$. Using this
isomorphism, a monoidal structure can be given to the identity
endofunctor over $\C$, giving a monoidal functor from $\C_{\oX}$ to
$\C_{{\oX}^{\op}}$.
\begin{mtheorem}
Let $\C_{\oX} = (\C, \oX, \oI, \alpha, \lambda, \rho, \gamma)$ be
a symmetric monoidal category, then we have a monoidal functor
 $(\Id, \gamma, \id) : \C_{\oX} \to\C_{{\oX}^{\op}}$.
\end{mtheorem}

If we apply Theorem~\ref{thm:monoid} to a monoid $M$ in $\C_{\oX}$, we
obtain a monoid in $\C_{{\oX}^{\op}}$. From
Theorem~\ref{thm:monoidop}, this monoid can be converted to a monoid
in $\C_{\oX}$. This last monoid is what we call the \emph{reversed
  monoid of $M$}.

As already mentioned, $\EndD$ is a symmetric monoidal category, 
and therefore the reverse monoid construction can be applied to a monoid
in $\EndD$. The resulting monoid is known as the \emph{reversed
applicative}~\cite{bird2013}.



\begin{coder}
The reversed applicative is implemented as:

\begin{hscode}\SaveRestoreHook
\column{B}{@{}>{\hspre}l<{\hspost}@{}}%
\column{3}{@{}>{\hspre}l<{\hspost}@{}}%
\column{20}{@{}>{\hspre}l<{\hspost}@{}}%
\column{E}{@{}>{\hspre}l<{\hspost}@{}}%
\>[B]{}\mathkw{data}\;\mathsf{Rev}\;\Varid{f}\;\Varid{x}\mathrel{=}\mathsf{Rev}\;(\Varid{f}\;\Varid{x})\;\mathkw{deriving}\;\mathsf{Functor}{}\<[E]%
\\[\blanklineskip]%
\>[B]{}\mathkw{instance}\;\mathsf{Applicative}\;\Varid{f}\Rightarrow \mathsf{Applicative}\;(\mathsf{Rev}\;\Varid{f})\;\mathkw{where}{}\<[E]%
\\
\>[B]{}\hsindent{3}{}\<[3]%
\>[3]{}\FN{pure}{}\<[20]%
\>[20]{}\mathrel{=}\mathsf{Rev}\circ\FN{pure}{}\<[E]%
\\
\>[B]{}\hsindent{3}{}\<[3]%
\>[3]{}\mathsf{Rev}\;\Varid{f}\circledast\mathsf{Rev}\;\Varid{x}{}\<[20]%
\>[20]{}\mathrel{=}\mathsf{Rev}\;(\FN{pure}\;(\FN{flip}\;(\mathbin{\$}))\circledast\Varid{x}\circledast\Varid{f}){}\<[E]%
\ColumnHook
\end{hscode}\resethooks
In intuitive terms, the difference between \ensuremath{\Varid{f}} and \ensuremath{\mathsf{Rev}\;\Varid{f}} as
applicative functors is that \ensuremath{\mathsf{Rev}\;\Varid{f}} sequences
the order of effects in the opposite order~\cite{bird2013}.

\end{coder}

%

\section{Conclusion}
\label{sec:conclusion}
We have seen how monads, applicative functors and arrows can be cast
as monoids in a monoidal category. We exploited this uniformity in
order to obtain free constructions and representations for the three
notions of computation. 
%
We have provided Haskell code for all of the concepts, showing that
the ideas can be readily implemented without difficulty. The
representations for applicative functors and arrows are new and they
optimise code in the same cases the codensity transformation and
difference lists work well: when the binary operation of the monoid is
expensive on its first argument and therefore, we want to associate a
sequence of computations to the right. In order to prove this
formally, we could adopt the framework of Hackett and
Hutton~\shortcite{HH2014}.

The constructions presented for monads are well
known~\cite{macLaneS:catwm}.  Day has shown the equivalence of lax
monoidal functors and monoids with respect to the Day
convolution~\cite{Day70}. However, in the functional programming
community, this fact is not well-known. The construction of free
applicatives is described by Capriotti and
Kaposi~\shortcite{Capriotti2014}. While they provide plenty of
motivation for the use of the free applicative functor, we give a
detailed description of its origin, as we arrive at it instantiating a
general description of free monoids to the category of endofunctors
which is monoidal with respect to the Day convolution.

There are several works analysing the formulation of arrows as
monoids~\cite{jacobs2009categorical,Atkey:2011,Asada10,AsadaH10}. We
differentiate from their work in our treatment of the strength. We
believe our approach leads to simpler definitions, as only standard
monoidal categories are used. Moreover, our definition of the free arrow
is possible thanks to this simpler approach.


Jaskelioff and Moggi~\shortcite{JM:TCS:2010} use the Cayley
representation for monoids in a monoidal category in order to lift
operations through monoid transformers. However, the only instances
considered are monads.

For simplicity, we analysed the above notions of computations as
$\Set$ functors. However, for size reasons, many constructions were
restricted to small functors, which are extensions of functors from
small categories. Alternatively, we could have worked with accessible
functors~\cite{Adamek:1994} (which are equivalent to small functors), or we could
have worked directly with functors from small categories, as it is
done in relative monads~\cite{relativemonads}. However, by working
with small functors the category theory is less heavy and the
implementation in Haskell is more direct.

In functional programming, for each of the three notions of computation
that we considered, there are variants which add structure. For
example, monads can be extended with \ensuremath{\mathsf{MonadPlus}}, applicative functors
with \ensuremath{\mathsf{Alternative}}, and arrows with \ensuremath{\mathsf{ArrowChoice}}, to name just a
few. It would be interesting to analyse the relation between the
different extensions from the point of view of monoidal categories
with extra structure.

Unifying different concepts under one common framework is a worthy goal
as it deepens our understanding and it allows us to relate, compare, and
translate ideas. It has long been recognised that category theory is
an ideal tool for this task~\cite{Reynolds80} and this article provides
a bit more evidence of it.

\subsection*{Acknowledgements}
We thank Ond\v{r}ej Ryp\'a\v{c}ek and Jennifer Hackett for their
insightful comments on an early version of this document. This work
was funded by the Agencia Nacional de Promoción Científica y
Tecnológica (PICT 2009--15) and Consejo Nacional de Investigaciones
Científicas y Técnicas (CONICET).
%

\bibliographystyle{jfp}
\bibliography{main}

\end{document}